\documentclass{raa_twocolumn}
\usepackage{graphicx,times}
\usepackage{natbib}
\usepackage{amssymb,amsmath,bm}
\usepackage{booktabs}
\usepackage[percent]{overpic}
\usepackage{microtype}
\usepackage{placeins}
\usepackage{needspace}
\usepackage{float}
\usepackage{stfloats}
\usepackage{hyperref}
\bibpunct{(}{)}{;}{a}{}{,}

\hypersetup{
  hidelinks,
  pdftitle={Inference of Cosmic Dawn and Reionization Parameters from the 21 cm Global Spectrum Measurement of the DSL Mission},
  pdfauthor={Chengliang Xu, Meng Zhou, Furen Deng, Yidong Xu, Xuelei Chen},
  pdfkeywords={dark ages, reionization, first stars; methods: statistical; radio lines: general; space vehicles: instruments}
}
\DeclareUnicodeCharacter{2011}{-}

\renewcommand{\makeheadbox}{%
  \hbox to\textwidth{%
    \begin{minipage}{0.75\textwidth}
      \small Research in Astronomy and Astrophysics\\
      \url{https://www.raa-journal.org/}
    \end{minipage}\hfill\RAAlogo}}

\begin{document}
\title{Inference of Cosmic Dawn and Reionization Parameters from the 21\,cm Global Spectrum Measurement of the DSL Mission}
\volnopage{}
\date{}
\setcounter{page}{1}
\author{Chengliang Xu\inst{1,2}
  \and Meng Zhou\inst{1}
  \and Furen Deng\inst{1}
  \and Yidong Xu\inst{1,\star}
  \and Xuelei Chen\inst{1,2,\star}}
\institute{State Key Laboratory of Radio Astronomy and Technology, National Astronomical Observatories, CAS, 20A Datun Road, Chaoyang District, Beijing 100101, People's Republic of China
  \and School of Astronomy and Space Science, University of Chinese Academy of Sciences, Beijing 100049, People's Republic of China}
\abstract{The 21\,cm global spectrum is an important observable for cosmic dawn and epoch of reionization. The \textit{Discovering the Sky at the Longest Wavelengths} mission (DSL, also known as Hongmeng), will measure the 21\,cm global spectrum from the lunar orbit, thus avoiding the systematic effects common in ground based experiments, potentially producing the most accurate measurement to date. In this paper, we present a Bayesian inference framework, and investigate the constraints that can be placed on the cosmic dawn and reionization model parameters, either from the 21~cm global spectrum data alone, or from its combination with 21\,cm spatial power-spectrum data. For the global spectrum data, we assume a residual error equivalent to the thermal noise of a total observation of 30, 300, or 3,000 orbit cycles of DSL. For the power spectrum, we adopt representative forecasts for HERA or SKA-Low.
We find that there is generally a degeneracy between the foreground and the 21~cm global signal, and one has to make some restrictions on the form of either the foreground or the 21~cm global spectrum to break this degeneracy. Using the physically motivated emulator \texttt{21cmEMU}, we generate the 21\,cm global spectrum and power spectrum under a common astrophysical parameterization, while marginalizing over the foreground parameters. We derive constraints for two representative models, with their global spectrum absorption minimum at $\nu=$105 and 94 MHz respectively. With only the global spectrum data, there remains much degeneracy in the astrophysical parameters. We find that the spatial power spectrum is highly complementary to the global spectrum, which can break the degeneracy and tighten up the constraints significantly.
}

\authorrunning{C. Xu et al.}
\titlerunning{Cosmic Dawn and Reionization Constraints from DSL}
\maketitle
\begingroup
\renewcommand{\thefootnote}{\ensuremath{\star}}
\footnotetext[1]{Corresponding authors (Yidong Xu, email:
\href{mailto:xuyd@nao.cas.cn}{xuyd@nao.cas.cn}; Xuelei Chen, email:
\href{mailto:xuelei@cosmology.bao.ac.cn}{xuelei@cosmology.bao.ac.cn}).}
\endgroup
\setcounter{footnote}{0}

\section{Introduction}

The \textit{Cosmic Dawn (CD)} and subsequent \textit{Epoch of Reionization (EoR)} mark the formation of the first luminous sources and the transition of the intergalactic medium (IGM) from neutral to ionized. The redshifted 21\,cm line provides a sensitive probe of these epochs, tracing both the astrophysical processes and thermal history that shape early structure formation \citep{Mesinger2011,Furlanetto2006,Pritchard2012,Bera2022,Shimabukuro2023,Fialkov2024}.

One of the primary observational methods for accessing the high-redshift 21\,cm signal is the measurement of the {\it power spectrum} of its spatial fluctuations. Experiments such as LOFAR \citep{Mertens2020} and HERA \citep{HERA2022} have already placed stringent upper limits on the 21\,cm power spectrum during reionization, while the forthcoming Square Kilometre Array (SKA) is expected to provide precision constraints on the timing and morphology of reionization, with its low-frequency component targeting the EoR and CD epochs \citep{Mellema2013,Koopmans2015}. Complementary to these efforts, measurements of the \textit{sky-averaged (global) spectrum} aim to detect the evolution of the mean brightness temperature as a function of frequency, which directly probes the absorption and emission features associated with the CD and the EoR. A number of instruments—including EDGES \citep{Bowman2018}, SCI-HI \citep{Voytek2014}, BIGHORNS \citep{Sokolowski2015}, LEDA \citep{Price2017,Bernardi2016}, PRIzM \citep{Philip2019}, SARAS \citep{Singh2022}, and REACH \citep{deLeraAcedo2022REACH}—have pursued this goal. The most widely discussed result to date is the EDGES Low-Band report of a flattened absorption trough at $\sim 78$\,MHz with an amplitude of $\sim 0.5$\,K. This signal is significantly deeper than expected from standard astrophysical models and may suggest either exotic physics or unexpected astrophysical processes \citep{Bowman2018,Barkana2018}. However, this claim has been contested, as the SARAS-3 experiment reported a non-detection that rules out the EDGES profile under broad conditions \citep{Singh2022}. Ongoing (e.g., REACH \citep{deLeraAcedo2022REACH}, PRIzM \citep{Philip2019}, LEDA \citep{Bernardi2016}) and planned efforts (e.g., \citealt{Sun2024, xu2025progress, zhu2021, xu2021, Sun:2025}) are dedicated to improving calibration and systematic control to provide independent tests of the global signal.

However, ground-based global 21\,cm experiments are subject to severe systematic limitations that complicate the extraction of the faint cosmological signal. The ionosphere introduces frequency-dependent refraction, absorption, and emission, with amplitudes often exceeding the expected 21\,cm signal, and its variability over time and position induces spectral distortions that do not simply average out \citep{Shen2021}. In addition, antenna-beam chromaticity couples bright astrophysical foregrounds into the frequency domain, generating artificial spectral structures that can mimic or obscure the global signal \citep{Vedantham2014,Wang2024}. Further complications arise from terrestrial radio-frequency interference (RFI), which is particularly problematic at the low frequencies ($< 200$\,MHz) relevant for Cosmic Dawn observations, and from reflections off the ground and nearby structures, all of which require extremely precise calibration and modeling. These challenges make it difficult for ground-based experiments alone to deliver robust detections of the global 21\,cm signal from the CD and the EoR.

Space-based and lunar-orbit platforms provide a promising avenue for overcoming these fundamental limitations. By operating above the Earth's atmosphere, they eliminate ionospheric distortions, and when placed on the lunar far side, they benefit from natural shielding against terrestrial RFI. Several mission concepts have been proposed to exploit these advantages. These include dedicated orbital missions and CubeSat pathfinders such as the Dark Ages Radio Explorer (DARE) \citep{Burns2012DARE}, the Dark Ages Polarimetry Pathfinder (DAPPER) \citep{Burns2018DAPPER}, the Netherlands–China Low-Frequency Explorer (NCLE) \citep{Karapakula2024NCLE}, as well as the Indian PRATUSH (Probing ReionizATion of the Universe using Signal from Hydrogen) mission \citep{Rao2025}, the SEAMS (Space Electric and Magnetic Sensors) concept \citep{Tanti2023}, and the lunar-orbiting CosmoCube \citep{Artuc2024}. On the lunar surface, ambitious projects target the CD and Dark Ages with high-sensitivity arrays; these include FARSIDE \citep{Burns2021}, the Lunar Crater Radio Telescope (LCRT) \citep{Bandyopadhyay2021}, the Large-scale Array for Radio Astronomy on the Farside (LARAF) \citep{Chen2024}, and the Astronomical Lunar Observatory (ALO) concept \citep{KleinWolt2012}. LuSEE-Night \citep{Bale2023} represents a critical immediate step, serving as a pathfinder instrument scheduled for lunar deployment to test these low-frequency technologies in a radio-quiet environment.

Among these, the \textit{Discovering the Sky at the Longest Wavelengths} mission (DSL, also known by its Chinese name \textit{Hongmeng} mission) stands out as one of the most powerful yet relatively mature designs: a constellation of small lunar-orbiting satellites, forming an array for interferometric imaging and global-spectrum measurement in the 0.1--30\,MHz band, and one satellite dedicated to measuring the global 21\,cm spectrum in the 30--120\,MHz band \citep{Chen2020,Chen2021DSL,Chen2023}. These are called the DSL low-frequency and high-frequency bands, respectively. In this paper, we focus on data from the high-frequency band, which targets the redshifted 21\,cm signal from the CD and the EoR. By operating on the lunar far side, DSL suppresses terrestrial radio-frequency interference and avoids several major ground-based systematics, including ionospheric absorption/refraction and ground reflections. Analyses show that DSL can measure the global spectrum with high precision, even in the presence of realistic instrumental and environmental systematics such as chromatic beam effects \citep{Shi2022,Liu2024,Ao2025,Li2025}.

However, the foreground which is orders of magnitude stronger than the redshifted 21~cm signal poses a great challenge. Considering the various physical origins of the foreground, such as the Galactic synchrotron radiation and free-free radiation that vary with different directions, the extragalactic radio sources of different spectra, and the free-free absorptions at low frequency by the interstellar medium, the foreground spectrum may not be a simple power law. There is then some degeneracy between the redshifted 21~cm global signal and the foreground, which requires some restrictions on either the foreground model or the 21\,cm signal.

In this work, we develop a Bayesian inference framework to extract information from the 21\,cm global spectrum data, either alone or in combination with the 21\,cm spatial power spectrum data, to infer the cosmic dawn and reionization parameters. Most previous studies have focused on spatial fluctuation statistics, for example \texttt{21CMMC} \citep{Greig2015}, the EOS simulation framework \citep{Mesinger2016}, and simulation-based inference (SBI) methods \citep{Cooper2025}. Direct inference from the global spectrum has received less attention, with only a few studies under simplified assumptions \citep{Gu2020,Bevins2021}. A common limitation for both is the large computational cost of inference over $\sim10$ or more parameters. A practical way to address this is to use emulators such as \texttt{Globalemu} \citep{Bevins2021} and \texttt{21cmVAE} \citep{Bye2021}, which provide fast forward modeling and enable efficient MCMC exploration. Using fast, physically motivated forward modeling with \texttt{21cmEMU} \citep{Breitman2023}, we generate astrophysical predictions for both observables within a common astrophysical parameter space, and pass them through experiment-specific models to simulate mock datasets for the DSL radiometer and the HERA/SKA-Low interferometers under two fiducial scenarios. 
We then use Markov chain Monte Carlo (MCMC) to jointly sample the foreground model and astrophysical parameters. Marginalizing over parameterized foregrounds, we derive the constraints on the astrophysical parameters, and show that combining global and power-spectrum information helps break the parameter degeneracies and further tightens constraints on the model parameters.

The paper is organized as follows: Section \ref{sec:simulation} details the simulation of the 21\,cm signal, foregrounds, and mock observations, and introduces our Bayesian inference framework. We present the parameter constraints from DSL alone and the joint analysis  with spatial power
spectrum experiments in Section \ref{sec:results}, and conclude with a discussion of our findings in Section \ref{sec:conclusion}.

\section{Simulation}\label{sec:simulation}

In this section, we describe the simulation of the 21\,cm signal and its foreground. We first briefly introduce the forward modeling from the \texttt{21cmEMU} emulator, which is based on simulation data generated by \texttt{21cmFAST}. Then we introduce the modeling of the foreground sky map. Finally, we detail how we generate the mock data for the 21\,cm global and spatial power spectra, with realistic experimental configurations.

\subsection{21 cm Signal Modeling}
\label{sec:21cm model}
\begin{figure}[tbp]
\centering
\includegraphics[width=0.95\columnwidth]{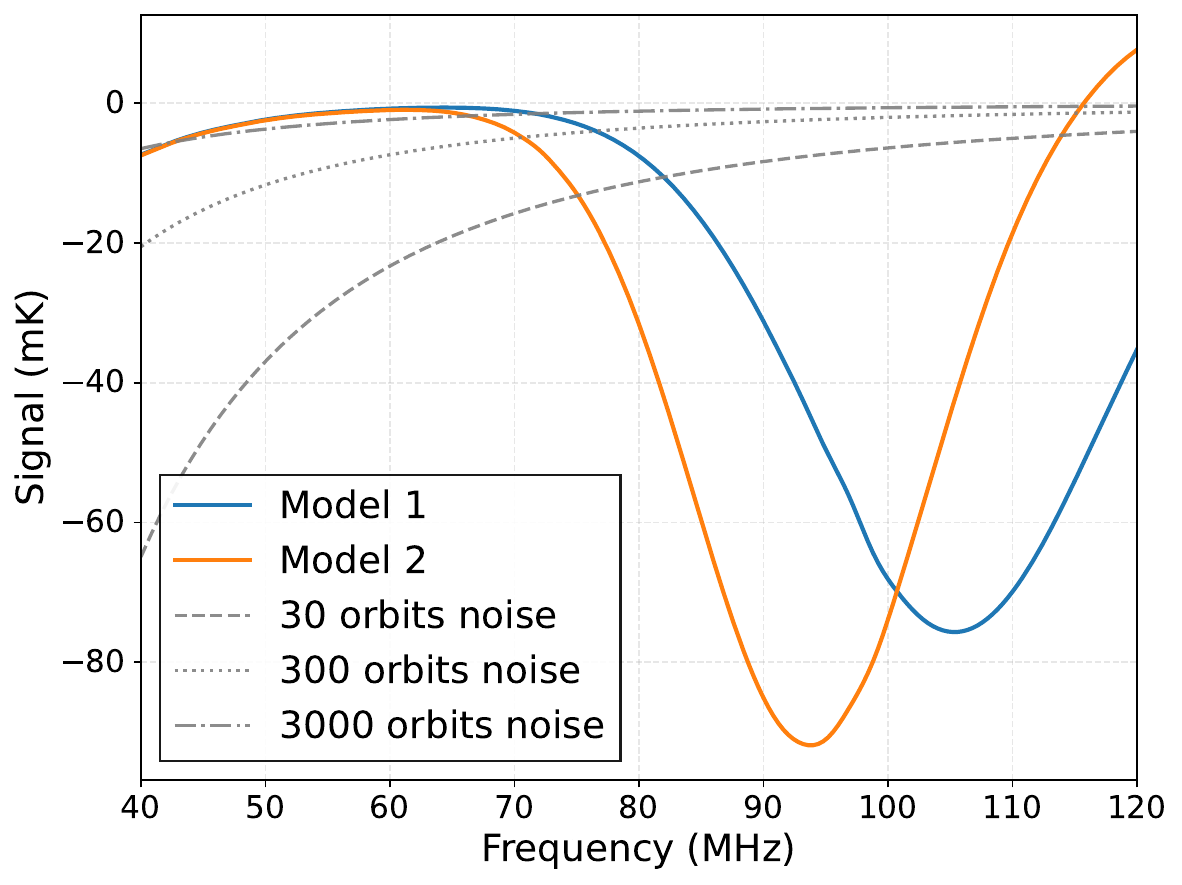}
\caption{The global 21\,cm spectra for two typical models: Model~1 (blue solid curve, $t_*=0.5$, with an absorption minimum of $\approx -76$\,mK at $\nu \approx 105$\,MHz) and Model~2 (orange solid curve, $t_*=0.1$, with a deeper minimum of $\approx -92$\,mK at $\nu \approx 94$\,MHz, crossing into emission at $\nu \approx 117$\,MHz), over the DSL high-frequency band. The grey curves show the negative thermal-noise levels for integrations of 30, 300, and 3,000 orbits (dashed, dotted, and dash-dotted curves, respectively).}
\label{fig:fiducial_signal_comparison}
\end{figure}
In addition to the cosmological model, the 21\,cm signal from cosmic reionization and CD should be governed by many astrophysical processes, including star formation (UV and Ly$\alpha$ photon production), ionizing photon escape, feedback on galaxy formation, and X‑ray heating. To efficiently sample astrophysical parameters that affect the 21\,cm signal $T_{21}(\nu; \boldsymbol{\theta}_{21})$, we employ a publicly available emulator, \texttt{21cmEMU} \citep{Breitman2023}, which is based on the summary observables of the semi-numerical simulation \texttt{21cmFASTv3} \citep{Mesinger2011,parkInferringAstrophysicsReionization2019,2020JOSS....5.2582M}. It is trained on $1.3 \times 10^6$ posterior samples from the inference in \citet{2022ApJ...924...51A}. It can output the 21\,cm global spectrum, mean neutral fraction, UV luminosity function, the mean spin temperature, and the 21\,cm power spectrum with a fractional error lower than 2.4\% and a computational cost reduced by a factor of $\sim 10^4$ compared to \texttt{21cmFAST}. This makes it practical to explore the high-dimensional astrophysical parameter space in Bayesian inference, and it has been validated and applied in various studies—ranging from EDGES global-signal reanalysis to HERA signal-loss modeling and posterior-recovery tests \citep{Cang2024EDGES,Pascua2024SignalLoss,Bevins2025PosteriorRecovery}.

\begin{table*}[tbp]
\centering
\caption{Astrophysical Parameters and Adopted Fiducial Values for Model~1 and Model~2, Together with the Prior Ranges for Inference.}
\label{tab:parameter}
{
\begin{tabular}{llllll}
\hline
Parameter & Model 1 & Model 2 & Units & Allowed range & Flat prior \\
\hline
$f_{\ast,10}$ & 0.05 & 0.05 & -- & 0.001--1 & log \\
$\alpha_{\ast}$ & 0.5 & 0.5 & -- & $-0.5$--1 & linear \\
$f_{\rm esc,10}$ & 0.1 & 0.1 & -- & 0.001--1 & log \\
$\alpha_{\rm esc}$ & $-0.5$ & $-0.5$ & -- & $-1$--0.5 & linear \\
$M_{\rm turn}$ & $5\times10^8$ & $5\times10^8$ & $M_\odot$ & $10^8$--$10^{10}$ & log \\
$t_{\ast}$ & 0.5 & 0.1 & -- & 0.01--1 & linear \\
$\dfrac{L_{\rm X<2\,keV}}{\rm SFR}$ & $10^{40.5}$ & $10^{40.5}$ & $\mathrm{erg\,s^{-1}\,M_\odot^{-1}\,yr}$ & $10^{38}$--$10^{42}$ & log \\
$E_0$ & 0.5 & 0.5 & keV & 0.1--1.5 & linear \\
$\alpha_{\rm X}$ & 1.0 & 1.0 & -- & $-1.0$--3.0 & linear \\
\hline
\end{tabular}}
\end{table*}

The relevant astrophysical processes in \texttt{21cmEMU} and \texttt{21cmFAST} are parameterized by nine parameters \citep{parkInferringAstrophysicsReionization2019,Breitman2023}, summarized with their fiducial values and prior ranges in Table~\ref{tab:parameter}. First, star formation is modeled via a power law in the stellar‑to‑halo mass relation (e.g., \citealt{Moster2013,Behroozi2013}),
\begin{equation}
    \frac{M_*}{M_h}=f_{*,10}\left(\frac{M_h}{10^{10}M_\odot}\right)^{\alpha_*}\frac{\Omega_b}{\Omega_m},
\end{equation}
where $M_h$ is the halo mass, $\Omega_b/\Omega_m$ is the cosmic baryon fraction, $f_{*,10}$ is the star-formation efficiency normalization at halo mass $M_h=10^{10}\,M_\odot$, and $\alpha_*$ is the power-law scaling index with halo mass.

Assuming that star formation occurs on a timescale proportional to the Hubble time $H(z)^{-1}$ \citep{Somerville1999,Cole2000,Barkana2001,Furlanetto2004,Mesinger2011}, the star formation rate (SFR) can be approximated by
\begin{equation}
    \dot M_* = \frac{M_*}{t_* H^{-1}(z)},
\end{equation}
where $t_*$ is a dimensionless star-formation timescale in units of the Hubble time, so that a smaller $t_*$ corresponds to a higher star-formation rate.
Given the total ionizing photon budget, the progress of reionization is also determined by the escape fraction of ionizing radiation \citep{parkInferringAstrophysicsReionization2019,Kimm2014,Paardekooper2015,Wise2014}, which is parameterized by
\begin{equation}
    f_{\rm esc}(M_h)=f_{\rm esc,10}\left(\frac{M_h}{10^{10}M_\odot}\right)^{\alpha_{\rm esc}},
\end{equation}
where $f_{\rm esc,10}$ and $\alpha_{\rm esc}$ are free parameters.

Star formation can be suppressed in small-mass halos by inefficient gas cooling and pressure smoothing, as well as stellar/photo-heating feedback
\citep{HuiGnedin1997,Okamoto2008,SobacchiMesinger2013,parkInferringAstrophysicsReionization2019}. The suppression factor can be described by an exponential duty-cycle law,
\begin{equation}
    f_{\rm duty} = {\rm exp}\left(-\frac{M_{{\rm turn}}}{M_{h}}\right),
\end{equation}
 where $M_{\rm turn}$ is the characteristic turnover halo mass below which star formation is exponentially suppressed due to inefficient cooling and feedback effects.

The specific X-ray luminosity, which heats the IGM, is also assumed to be a power law in the photon energy, $L_X\sim E^{-\alpha_X}$, where the index $\alpha_X$ is a free parameter \citep{Furlanetto2006,Pritchard2012,SobacchiMesinger2013,Fialkov2014}. It is normalized by the integrated soft X-ray band ($<2$ keV) luminosity per unit SFR \citep{Msinger2013,Fragos2013}:
\begin{equation}
    \frac{L_{\rm X<2\,keV}}{\rm SFR}=\int_{E_0}^{2\,\rm keV}\frac{L_X}{\rm SFR}\,dE,
\label{eq:lxsoft_def}
\end{equation}
where $E_0$ is the energy threshold for X-ray photons capable of escaping the host galaxy.

We consider two representative astrophysical models, denoted Model~1 and Model~2 as our fiducial models. Their adopted parameters and the inference priors are listed in Table~\ref{tab:parameter}, and both global spectra are shown in Figure~\ref{fig:fiducial_signal_comparison}. These two models share the same fiducial parameters except for the star-formation timescale, $t_*=0.5$ in Model~1 and $t_*=0.1$ in Model~2. Model~1 closely follows the default configuration of Park et al.~\citep{parkInferringAstrophysicsReionization2019} and yields a relatively late and extended reionization history: a moderate star-formation efficiency, a relatively long star-formation timescale, and standard soft X-ray luminosity, together delay IGM heating and produce a broad global 21\,cm absorption trough with its minimum located at $\nu \approx 105$\,MHz ($z \approx 12.5$) and a depth of $\approx -76$\,mK. In the original \texttt{21cmEMU}/\texttt{21cmFAST} analysis, this model has a reionization midpoint at $z_{\rm re}\simeq 7.8$ with an effective duration of $\Delta z \sim 2$--$3$, broadly consistent with current CMB and quasar constraints. By contrast, Model~2 assumes a substantially shorter star-formation timescale ($t_*=0.1$), shifting the absorption trough to earlier times ($\nu \approx 94$\,MHz, $z \approx 14.1$, with a deeper minimum of $\approx -92$\,mK).

Altogether, these nine parameters—$\log_{10}f_{*,10}$, $\alpha_*$, $t_*$, $\log_{10}f_{\rm esc,10}$, $\alpha_{\rm esc}$, $\log_{10}M_{\rm turn}$, $\alpha_X$, $\log_{10}L_{\rm X<2\,\rm keV}/{\rm SFR}$, and $E_0$—encode the astrophysical processes of UV/Ly$\alpha$ coupling, ionization, and X-ray heating that sculpt the timing, amplitude, and morphology of the 21\,cm signal during Cosmic Dawn and reionization.

\begin{figure*}[tp]
  \centering
  \includegraphics[width=0.90\textwidth]{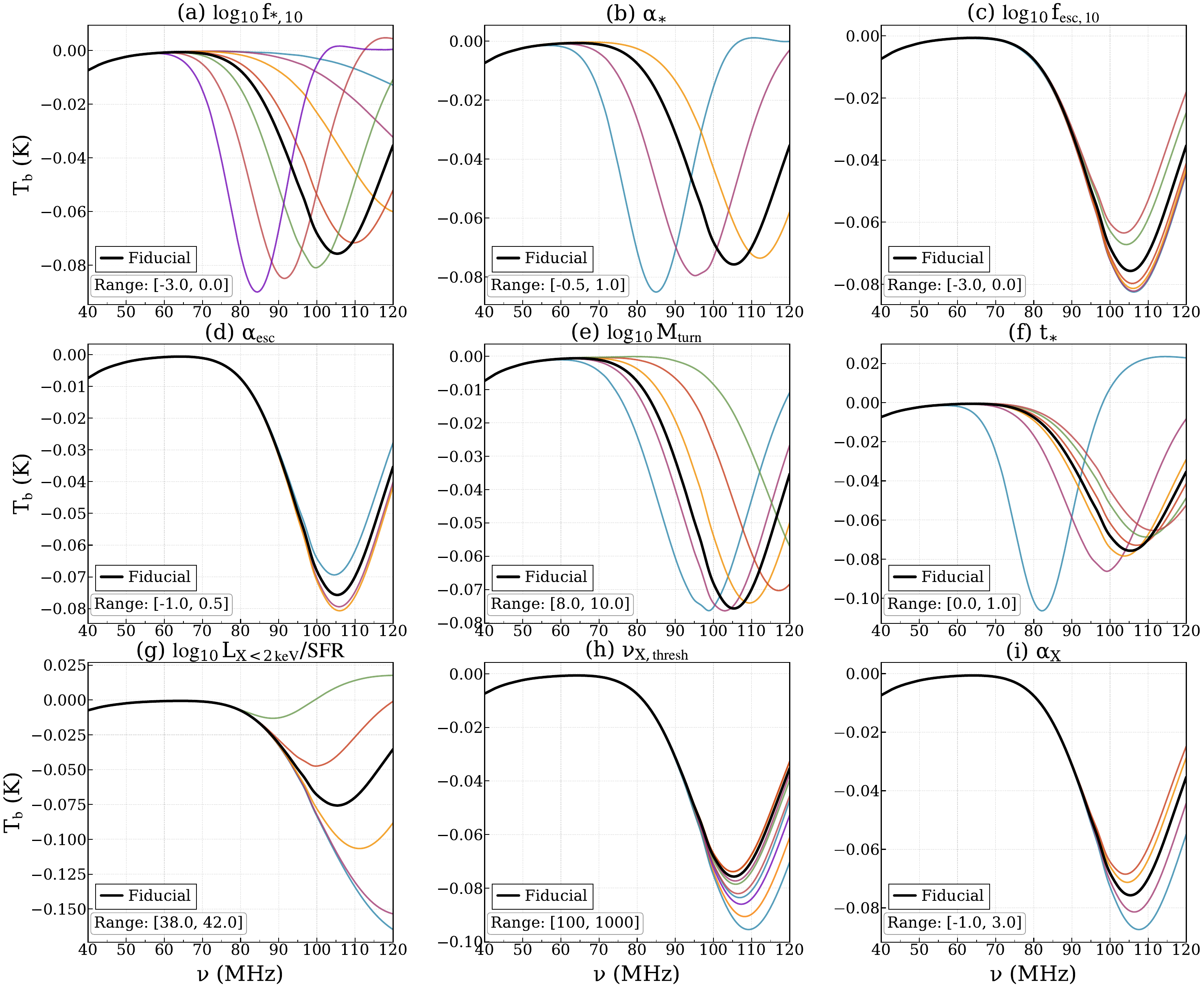}
  \caption{The global-spectrum response to parameter variations across the full parameter space of \texttt{21cmEMU}, evaluated around Model~1 as an illustrative fiducial case.}
  \label{fig:FullParamRange}
\end{figure*}

The 21\,cm signal is not equally sensitive to all nine parameters. Fixing some of them can simplify the inference problem and lower computational cost, without losing vital features of the modeling. We perform a parameter sensitivity analysis with respect to the 21\,cm global spectrum. In each panel of Figure~\ref{fig:FullParamRange}, we vary one parameter over a few values across the default range of the \texttt{21cmEMU} prior while keeping the other parameters fixed to their Model~1 fiducial values in Table~\ref{tab:parameter}, and show the resulting global 21\,cm spectra. We observe that five of the parameters, namely $\log_{10} f_{*,10}$, $\alpha_\ast$, $\log_{10} M_{\rm turn}$, $t_{\ast}$, and $\log_{10} \left(L_{\rm X<2\,\mathrm{keV}}/\mathrm{SFR}\right)$, exhibit significant influence on the shape of the global 21\,cm spectrum. As seen in Figure~\ref{fig:FullParamRange}, shortening the star-formation timescale $t_*$ accelerates both Ly$\alpha$ coupling and X-ray heating, noticeably advancing the absorption trough to lower frequencies (earlier times); variations in $f_{*,10}$ and $L_{\rm X<2\,\mathrm{keV}}/\mathrm{SFR}$ primarily govern the depth of the absorption feature and the timing of the transition to emission, while $\alpha_*$ and $M_{\rm turn}$ determine the mass dependence and suppression of star formation in low-mass halos. Therefore, we select these five parameters for our active inference in this work, while keeping the remaining four parameters ($f_{\rm esc,10}$, $\alpha_{\rm esc}$, $E_0$, and $\alpha_X$) fixed at their fiducial values in Table~\ref{tab:parameter}.

\subsection{The Foreground and Beam Response}
\label{subsec:fg_sky_map}

In the frequency range relevant for this work, $\nu \simeq 30$--$120\,\mathrm{MHz}$, the dominant foreground is diffuse Galactic synchrotron emission, which has a brightness temperature exceeding the cosmological 21\,cm signal by a few orders of magnitude. To better capture the low-frequency behavior of the foregrounds, we adopt the Ultra-long Wavelength Sky Model with Absorption (ULSA; \citealt{Cong2021ULSA}) as our diffuse sky model. ULSA is an observation-based Galactic radio sky model developed for the ultra-long wavelength regime and constructed to be valid down to frequencies of order $\sim 1$ MHz. It combines multi-frequency radio maps in the 35--80\,MHz band (from various low-frequency sky surveys) and the Haslam 408\,MHz map \citep{Haslam1982} with a three-dimensional model of the free-electron distribution (e.g., the NE2001 model; \citealt{Cordes2002NE2001I,Cordes2003NE2001II}) to predict the frequency-dependent brightness-temperature maps that naturally include the effect of Galactic free-free absorption. This physically motivated treatment predicts distinctive ultra-long-wavelength features---such as a comparatively dark Galactic plane, prominent absorption shadows associated with spiral arms, and a global spectral turnover at a few megahertz---that are not present in simple power-law extrapolations of higher-frequency data.

\begin{figure}[tbp]
\centering
\includegraphics[width=0.90\columnwidth]{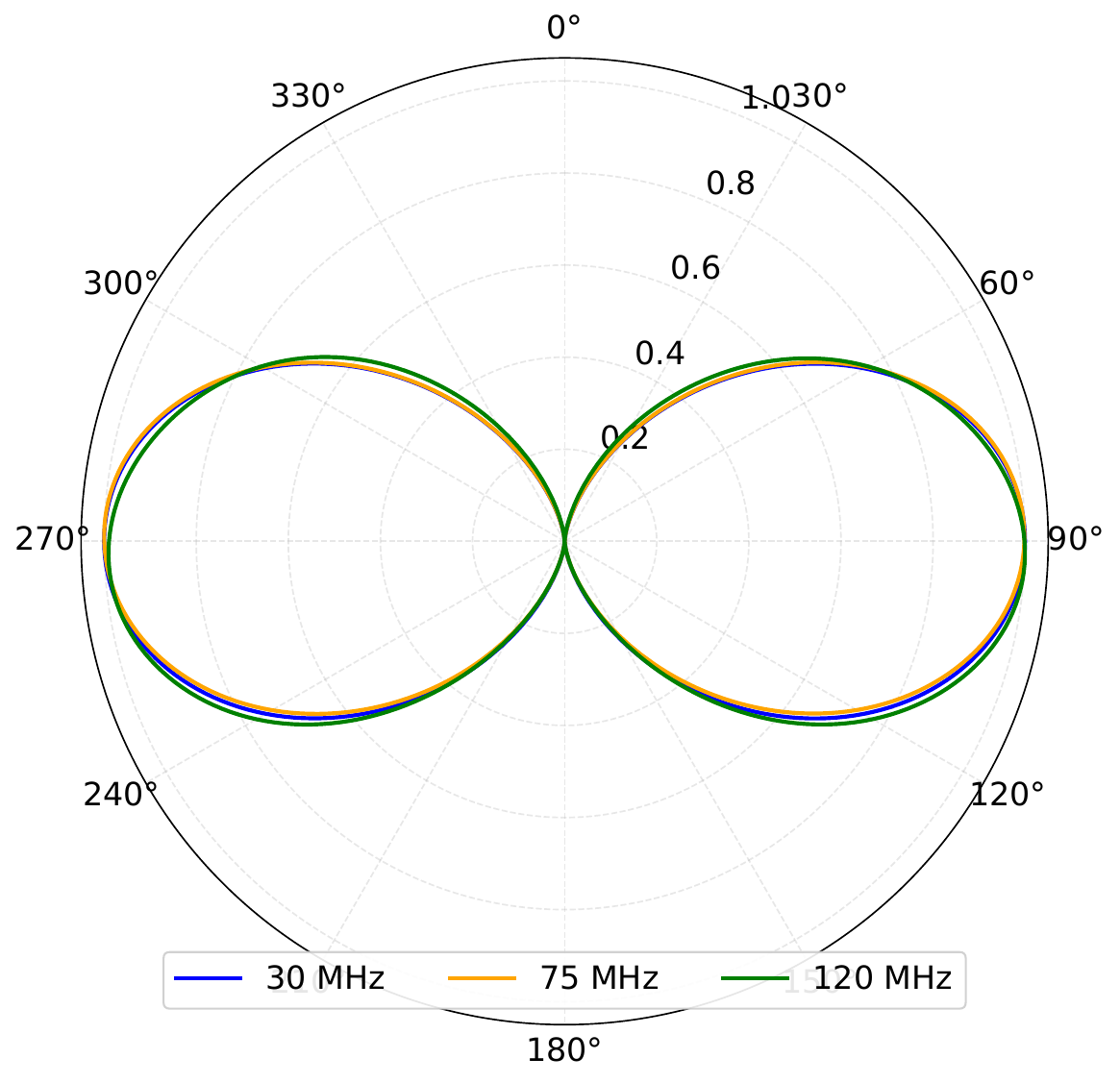}
\par\vspace{2pt}
\includegraphics[width=0.90\columnwidth]{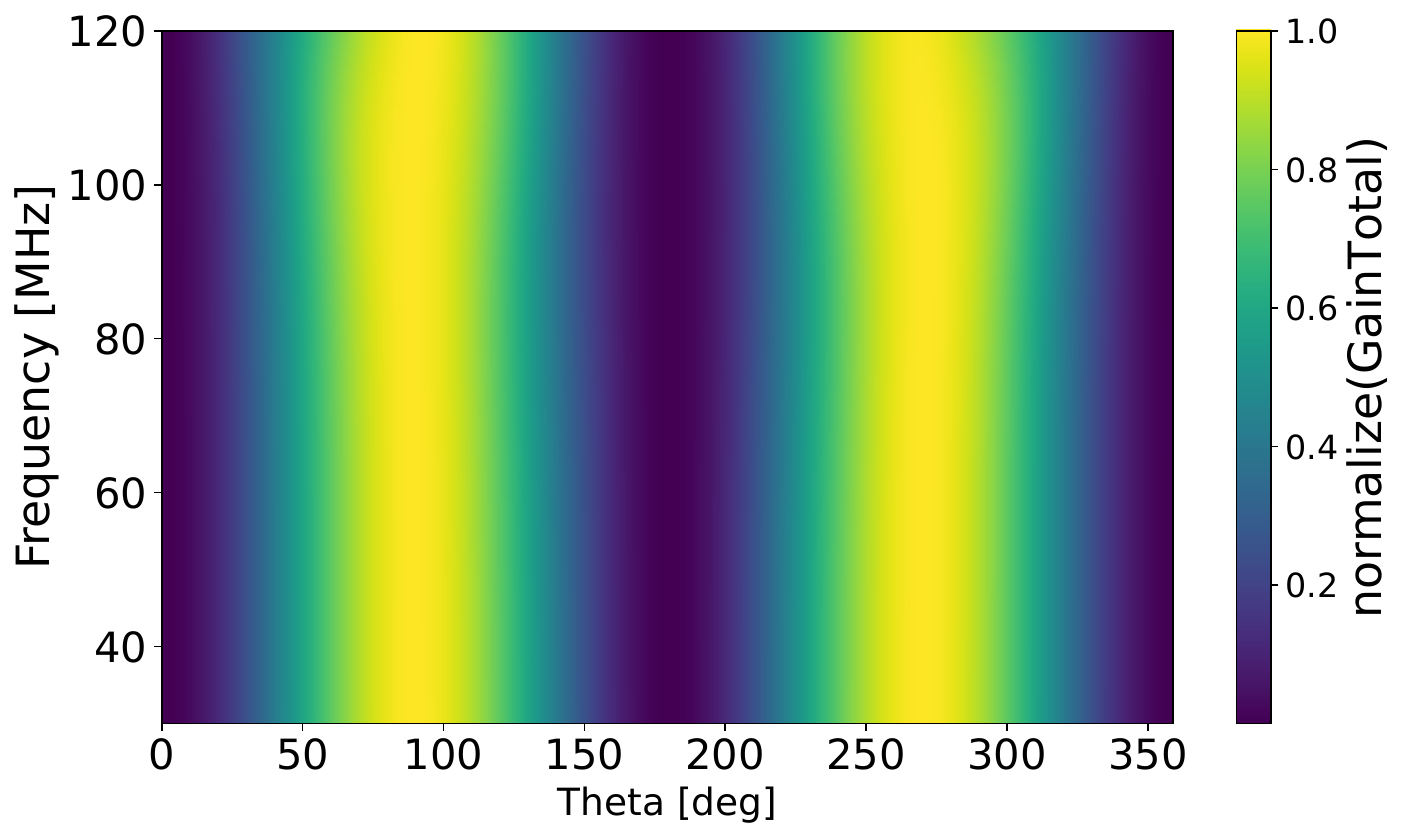}
\par
\caption{Simulated CB-1 beam pattern. Top: the polar projection of the CB-1 beam at 30, 75, and 120\,MHz; Bottom: the CB-1 primary beam as a function of frequency.}
\label{fig:cbbeam1}
\end{figure}

We model the foreground sky temperature with a logarithmic polynomial:
\begin{equation}
    T_{\mathrm{FG}}(\nu; \boldsymbol{\theta}_{\mathrm{FG}}) = \left(\frac{\nu}{\nu_{\mathrm{c}}}\right)^{-2.5} \sum_{n=0}^{N_{\mathrm{poly}}} a_n \left[\log \left(\frac{\nu}{\nu_{\mathrm{c}}}\right)\right]^n,
    \label{eq:fg_model}
\end{equation}
where $\nu$ is the observation frequency and $\nu_{\mathrm{c}} = 80$\,MHz serves as the reference frequency. The foreground parameter set $\boldsymbol{\theta}_{\mathrm{FG}} = \{a_n\}$ contains the polynomial coefficients to be inferred. Through empirical testing, we find that a seven-term logarithmic polynomial is required to achieve a foreground-subtraction residual smaller than the thermal noise for the 3{,}000-orbit case.

Antenna-beam chromaticity can introduce spectral distortions that bias signal recovery \citep{Shi2022}. We consider two beam models in this work to assess the impact of beam chromaticity on parameter inference.

First, we consider an ideal frequency-independent dipole beam, i.e., the beam pattern remains constant across all frequencies in the observational band. It provides a useful benchmark for isolating the effects of beam chromaticity from other systematic uncertainties.

Second, we employ a realistic beam model derived from detailed electromagnetic simulations of the so-called ``ice-cream'' antenna~\citep{Shi2022,Liu2024}.
This beam pattern exhibits a very small but non-zero frequency dependence across the observational band, particularly toward the edges of the band.
This design represents the expected performance of a realistic model of the DSL antenna, though the DSL antenna design is still evolving.
Hereafter, we refer to this model as ``CB-1'' (Chromatic Beam 1), as shown in Figure~\ref{fig:cbbeam1}. This is a deliberately mild chromaticity test case: the beam remains smooth overall, but retains enough frequency dependence to probe beam-induced parameter bias.
Our goal is to assess how these subtle chromatic effects, when coupled with thermal noise and foreground-modeling uncertainties, affect the scientific analysis and parameter recovery.

\subsection{DSL Global Spectrum Setup}

\begin{table}
\centering
\caption{DSL Global-Spectrum Measurement Parameters}
\label{tab:DSL_param}
\begin{tabular}{ll}
\hline
Parameter & Value \\
\hline
Orbit Altitude & 300\,km\\
Orbit Inclination & $30^\circ$\\
Orbit Period & 8249.6\,s\\
Far-side Time Fraction & 31\% \\
Precession Period & 1.3\,years \\
\hline
\end{tabular}
\end{table}

For the 21\,cm global-spectrum measurement by the DSL mission, we simulate the observation by following the orbital dynamics and sky coverage of the satellite. The DSL satellite operates in a lunar orbit at an altitude of approximately 300\,km, with an orbital period of $\sim 2.29$\,hours (8249.6\,s), an orbital inclination of $30^\circ$, and a precession period of $1.3$ years \citep{Chen2021DSL}. The antenna beam null is always pointed toward the lunar center. The observing time comprises those periods when the satellite is on the far side of the Moon, which account for approximately 31\% of each orbital period. These parameters are summarized in Table \ref{tab:DSL_param}.

Depending on the satellite's position, the Moon blocks part of the sky. We treat the Moon as an opaque sphere, as lunar radiation is relatively weak at these frequencies, and reflected emission is randomized, which introduces some additional noise but is unlikely to imprint any particular pattern \citep{Deng2025}.

The antenna temperature
$T_{\rm ant}(\nu_i, t_j)$ is given by
\begin{equation}
    \begin{split}
    T_{\rm ant}(\nu, t) =& \int d\Omega\, S\bigl(\hat{\mathbf{n}}, t\bigr)\,\\
    & B\bigl(\nu, \hat{\mathbf{n}}, t\bigr)\,
    \left[T_{\rm sky}(\nu, \hat{\mathbf{n}})+T_{\rm 21}(\nu)\right]
    \end{split}
\end{equation}
where $\hat{\mathbf{n}}$ denotes the sky direction, $S(\hat{\mathbf{n}}, t)$ is the visibility mask describing the lunar occultation (1 for visible sky, 0 for blocked), $B(\nu, \hat{\mathbf{n}}, t)$ is the primary beam response, $T_{\rm sky}$ is the foreground sky map generated by the ULSA model, and $T_{\rm 21}$ is the global 21\,cm spectrum generated by \texttt{21cmEMU}. We discretize the sky using the \textsc{HEALPix} scheme and set $N_{\mathrm{side}} = 64$, corresponding to a resolution of approximately $1^\circ$, which is sufficient for our global-spectrum model.

In this work, we first generate time-ordered data with an integration time of 8\,s and then average over multiple orbits to obtain a single time-averaged global spectrum as the observable. In principle, the time-dependent global spectrum could be used as the observable to infer more about the foreground sky and obtain a more robust global spectrum \citep{Ao2025, Li2025}. The averaged global spectrum is
\begin{equation}
    T_{\rm mock}(\nu_i) = \frac{1}{N_t}\sum_{j=1}^{N_t} \bigl[ T_{\rm ant}(\nu_i, t_j)+ n(\nu_i, t_j)\bigr]
    \label{eq:mock_obs}
\end{equation}
where $N_t$ is the total number of time samples and $n(\nu_i, t_j)$ is the thermal noise. Here we neglect variations in the system thermal noise caused by the slight variation of antenna temperature.

Although the observing band of the DSL high-frequency spectrometer is 30--120\,MHz, in this work we limit our analysis to 40--120\,MHz.  Our theoretical signal model (\texttt{21cmEMU}) has an effective high-redshift training limit of $z=35$, corresponding to $\approx 39.4$\,MHz, making predictions at frequencies below 40\,MHz unreliable. In any case, the Galactic synchrotron foreground noise scales as $T_{\rm sky} \sim \nu^{-2.5}$, meaning the 30--40\,MHz band suffers substantially higher noise while contributing negligibly to the parameter constraints.

For the global-spectrum measurement in the frequency band considered here, thermal noise is quickly suppressed with moderate integration time. However, systematic errors from calibration or beam chromaticity may set an error floor that does not decrease with longer integration. Their magnitude remains uncertain. We therefore consider a total of 30, 300, and 3,000 orbits, which, accounting for the actual observation time when the satellite is shielded from the Earth by the Moon, are equivalent to about 10, 100, and 1000 orbital periods of integration, with the global-spectrum uncertainty set by the corresponding thermal-noise level. Figure~\ref{fig:fiducial_signal_comparison} shows these three noise levels; the broad absorption features of both fiducial models are cleanly separable from the nominal noise floor only in deeper integrations.

\subsection{Power Spectrum Measurement Setup}

In addition to the DSL global mock data, we make forecasts of 21\,cm power-spectrum for interferometric experiments using the public sensitivity tool \texttt{21cmSense}\footnote{ \url{https://github.com/rasg-affiliates/21cmSense}.} \citep{pober2013,pober2014next,Liu2020}.

To generate the mock data for a given set of astrophysical parameters, we first evaluate the three-dimensional 21\,cm power spectrum using \texttt{21cmEMU} \citep{Breitman2023}. This signal is then projected onto the observational space using \texttt{21cmSense}, which computes the corresponding thermal-noise and sample-variance uncertainties in cylindrical $(k_\perp,k_\parallel)$ space, taking into account the array layout, primary beam, finite bandwidth, and $uv$ sampling.
We evaluate the sensitivity for each $uv$-bin of the interferometer array baseline distribution. The resulting forecast uncertainties account for thermal noise and sample variance and enter the power-spectrum likelihood defined below.

We consider two representative interferometric configurations: the Hydrogen Epoch of Reionization Array (HERA) and the SKA-Low Phase~1 AA4 layout. For HERA, we adopt a 331-element hexagonal core, a total integration time of $\sim 1400$\,h distributed over 67.4 observing days, and seven 20\,MHz bands with the central frequencies ranging from 60 to 200\,MHz \citep{pober2014next}.
For SKA-Low AA4, we assume 512 stations observing for 1080\,h (180 days with 6\,h per day) in nine 8\,MHz bands with the central frequencies ranging from 60 to 215\,MHz, with a maximum baseline cut of 1000\,m \citep{2019arXiv191212699B}.  The sky-temperature prescriptions and receiver temperatures are listed in Table~\ref{tab:psonly_obs_params}.

For both HERA and SKA-Low AA4, we adopt the ``Moderate'' foreground-avoidance prescription \citep{pober2013,pober2014next}, where the modes satisfying the condition
\begin{align}
    k_\parallel &\le \frac{H(z)D_c}{c(1+z)} k_\perp + k_{\rm buffer},
\end{align}
are excluded from the analysis. We set $k_{\rm buffer}=0.1\,h\,{\rm Mpc}^{-1}$.  Both forecasts use this prescription, and their respective array configurations and observational parameters are listed in Table~\ref{tab:psonly_obs_params}.
For each array, we generate mock power spectra for both Model~1 and Model~2 using their respective fiducial parameters, adopting the same Moderate foreground prescription.

\begin{table*}[tbp]
\centering
\caption{The Observational and Instrumental Parameters for the HERA \citep{pober2014next} and SKA-Low AA4 \citep{2019arXiv191212699B} Power-Spectrum Forecasts.\label{tab:psonly_obs_params} The SKA-Low AA4 forecast assumes coherent observations of a single deep field. Frequency ranges and band counts refer to the full forecast set; only bands within the adopted redshift interval enter the likelihood.
}
{\renewcommand{\baselinestretch}{1.0}
\footnotesize
\renewcommand{\arraystretch}{1.08}
\setlength{\tabcolsep}{3pt}
\begin{tabular*}{\textwidth}{@{\extracolsep{\fill}}llcc@{}}
\toprule
Parameter & Description & SKA-Low AA4 & HERA \\
\midrule
$N_{\mathrm{ants/stations}}$ & Number of array elements & 512 & 331 \\
$b_{\max}$ & Maximum baseline length & 1000\,m 
& 280\,m
\\[1pt]
$N_{\mathrm{days,eff}}$ & Effective number of days & 180 & 67.4 \\
$t_{\mathrm{day}}$ & Daily observing time & 6\,h & 21\,h \\
Total Time & Total integration time & 1080\,h & 1415.4\,h \\
$\tau_{\mathrm{int}}$ & Coherent integration (LST bin size) & 6\,h & 300\,s \\[1pt]
Forecast bands & \shortstack[l]{Band-center frequency range} & 60--215\,MHz (9 bands) & 60--200\,MHz (7 bands) \\[1pt]
$B$ & Nominal bandwidth per band & 8\,MHz & 20\,MHz \\
$N_{\mathrm{channels}}$ & Channels per band & 82 & 164 \\
$\Delta\nu$ & Channel width & $97.56$\,kHz & $122.07$\,kHz \\
$B_{\mathrm{eff}}$ & Effective bandwidth per band & 8\,MHz & 11.51\,MHz
\\[1pt]
$T_{\mathrm{sky}}$ & Sky temperature model & $150(\nu/150\,\mathrm{MHz})^{-2.5}$\,K & $150(\nu/150\,\mathrm{MHz})^{-2.5}$\,K \\
$T_{\mathrm{rcv}}$ & Receiver temperature & 100\,K (const.) & 600\,K@50\,MHz $\to$ 60\,K@200\,MHz \\[1pt]
Cosmology & Parameter set & Planck15 & Planck15 \\[1pt]
\bottomrule
\end{tabular*}
}
\end{table*}

\subsection{Method of Inference}
\label{sec:methodology}

We employ a Bayesian inference framework to recover the astrophysical parameters from the mock observations.
The two experiments target different observables, so we treat their likelihoods as independent and estimate them separately. The joint posterior is then proportional to the product of the two likelihood terms.
We generate representative samples of the combined constraints via importance sampling \citep{Gelman2013}, which effectively restricts the parameter space to regions consistent with both experiments.

The likelihood functions are $\mathcal{L} \propto \exp(-\chi^2/2)$. For the global-spectrum analysis for the DSL mission,
\begin{equation}
    \chi^2_{\rm global}(\boldsymbol{\theta}) = \sum_{n=1}^{N_{\mathrm{chan}}} \left[ \frac{ T_{\mathrm{mock}}(\nu_n) - T_{\mathrm{model}}(\nu_n; \boldsymbol{\theta}) }{\sigma_T(\nu_n)}\right]^2,
    \label{eq:chi2_global}
\end{equation}
where $T_{\rm mock}$ represents the mock observational data, $T_{\rm model}$ is the forward-modeled sky temperature including foregrounds and the 21\,cm signal,
$T_{\rm model}(\nu_i; \boldsymbol{\theta}) = T_{\mathrm{FG}}(\nu_i; \boldsymbol{\theta}_{\mathrm{FG}}) + T_{\rm 21}(\nu_i; \boldsymbol{\theta}_{\mathrm{21}})
$,
and $\sigma_T(\nu_n)$ denotes the noise level in the global spectrum.

For the 21\,cm spatial-fluctuation power spectrum from either the HERA or SKA observations,
\begin{equation}
    \begin{split}
    \chi^2_{\rm PS}(\boldsymbol{\theta}) =& \sum_{i=1}^{N_k}\sum_{j=1}^{N_{z}}\\
    &\left[\frac{ \Delta^2_{\mathrm{mock}}(k_i,z_j) -\Delta^2_{\mathrm{model}}(k_i,z_j; \boldsymbol{\theta}) }{\sigma_{\Delta^2}(k_i,z_j)}\right]^2.
    \end{split}
    \label{eq:chi2_ps}
\end{equation}
$\Delta^2$ is the dimensionless power spectrum. The uncertainty term $\sigma_{\Delta^2}$ includes both thermal noise and sample variance. Table~\ref{tab:psonly_obs_params} lists the ranges of the discrete forecast band-centre frequencies
and the individual bandwidths. After applying the adopted power-spectrum redshift cut, the likelihood uses six of the seven HERA bands, centred at 80, 120, 140, 160, 180, and 200\,MHz, and seven of the nine SKA bands, at the same centres plus 125\,MHz. These retained bands span $z\simeq6.1$--$16.8$, but the ranges of their centres do not imply continuous frequency coverage.

We assume the priors are either linearly uniform or logarithmically uniform, as detailed in Table~\ref{tab:parameter}. The five active astrophysical parameters $\boldsymbol{\theta}_{21}$, together with the foreground parameters $\boldsymbol{\theta}_{\mathrm{FG}}$, constitute the full parameter space $\boldsymbol{\theta} = \{\boldsymbol{\theta}_{\mathrm{FG}}, \boldsymbol{\theta}_{21}\}$ explored using the affine-invariant ensemble sampler \texttt{emcee} \citep{emcee}, where $\boldsymbol{\theta}_{21} = \{\log_{10}f_{*,10},\,\allowbreak \alpha_*,\,\allowbreak \log_{10}M_{\rm turn},\,\allowbreak t_*,\,\allowbreak \log_{10}(L_{\rm X<2\,\rm keV}/{\rm SFR})\}$. We run separate chains for the global and power-spectrum analyses before combining their posteriors to obtain final constraints.

\section{Results}\label{sec:results}
We now present the results of our simulations. We first consider an oversimplified case in which the 21\,cm signal can be inferred by imposing restrictions on the foreground. Unfortunately, this is generally not possible because the foreground is degenerate with the 21\,cm signal. We then consider a case in which the foreground is allowed to take a fairly general form, but the 21\,cm signal is restricted to spectra generated by the {\tt 21cmEMU} model; we place constraints on the astrophysical parameters of that model.

\subsection{Degeneracy with Foreground}\label{subsec:simple_fg}

For each fiducial model, we jointly fit the seven foreground parameters and the channel-wise signal values $T_{21,i}$ at a frequency resolution of 0.4\,MHz, assigning each $T_{21,i}$ a uniform prior over $[-1.0,0.5]$\,K. The foreground parameter fiducial values and two prior ranges used in the fits are listed in Table~\ref{tab:fg_prior_and_band}.

\begin{figure*}[tp]
\centering
\begin{minipage}[tbp]{0.48\textwidth}
\vspace{0pt}\centering
\includegraphics[width=\linewidth]{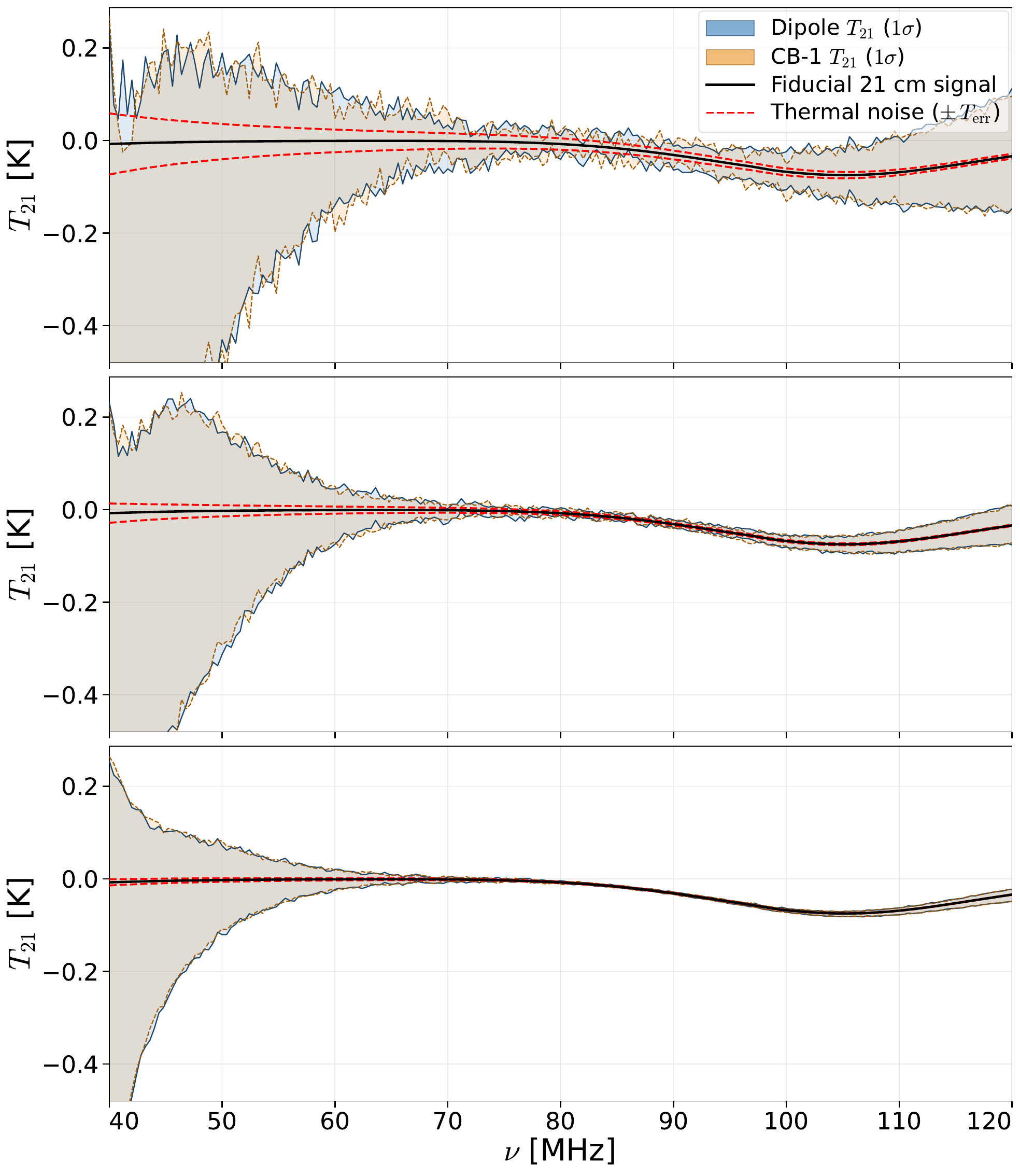}
\par\smallskip{\footnotesize{(a) Model 1}}
\end{minipage}
\hfill
\begin{minipage}[tbp]{0.48\textwidth}
\vspace{0pt}\centering
\includegraphics[width=\linewidth]{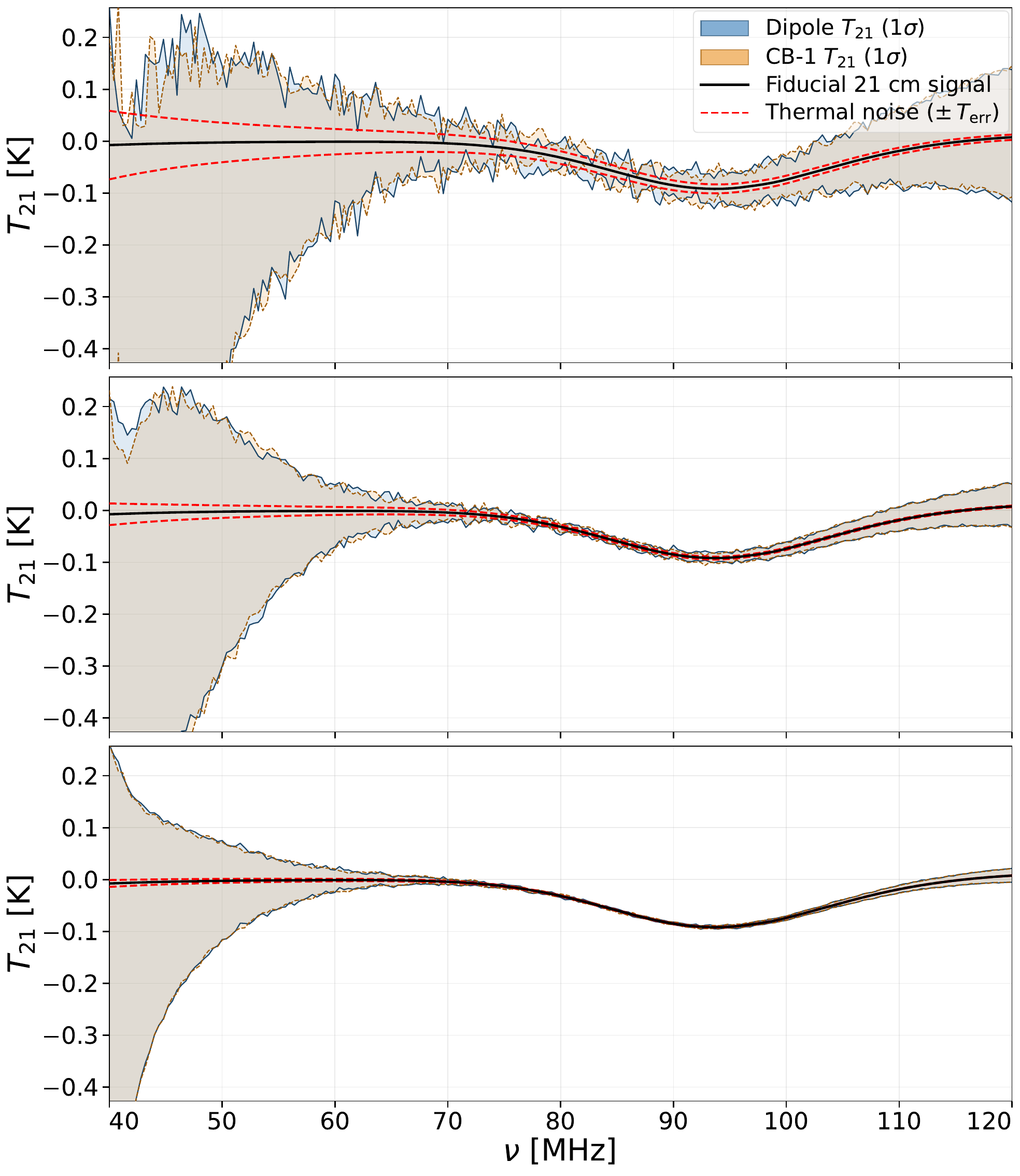}
\par\smallskip{\footnotesize{(b) Model 2}}
\end{minipage}
\caption{ Channel-wise recovered global-spectrum  with the ideal dipole and chromatic CB-1 beams for (a) Model~1 (left panel) and (b) Model~2 (right panel), adopting the tight foreground prior. From top to bottom, the three rows correspond to total observations of 30, 300, and 3,000 orbits. The black solid curve is the fiducial signal and the red dashed curves indicate the thermal-noise envelope ($\pm 1\sigma$). Blue solid and orange dash-dotted curves outline the $1\sigma$ posterior envelopes for the dipole and CB-1 analyses, respectively.}
\label{fig:global-spectrum-mock}
\end{figure*}

\begin{figure}[tbp]
\centering
\includegraphics[width=\columnwidth]{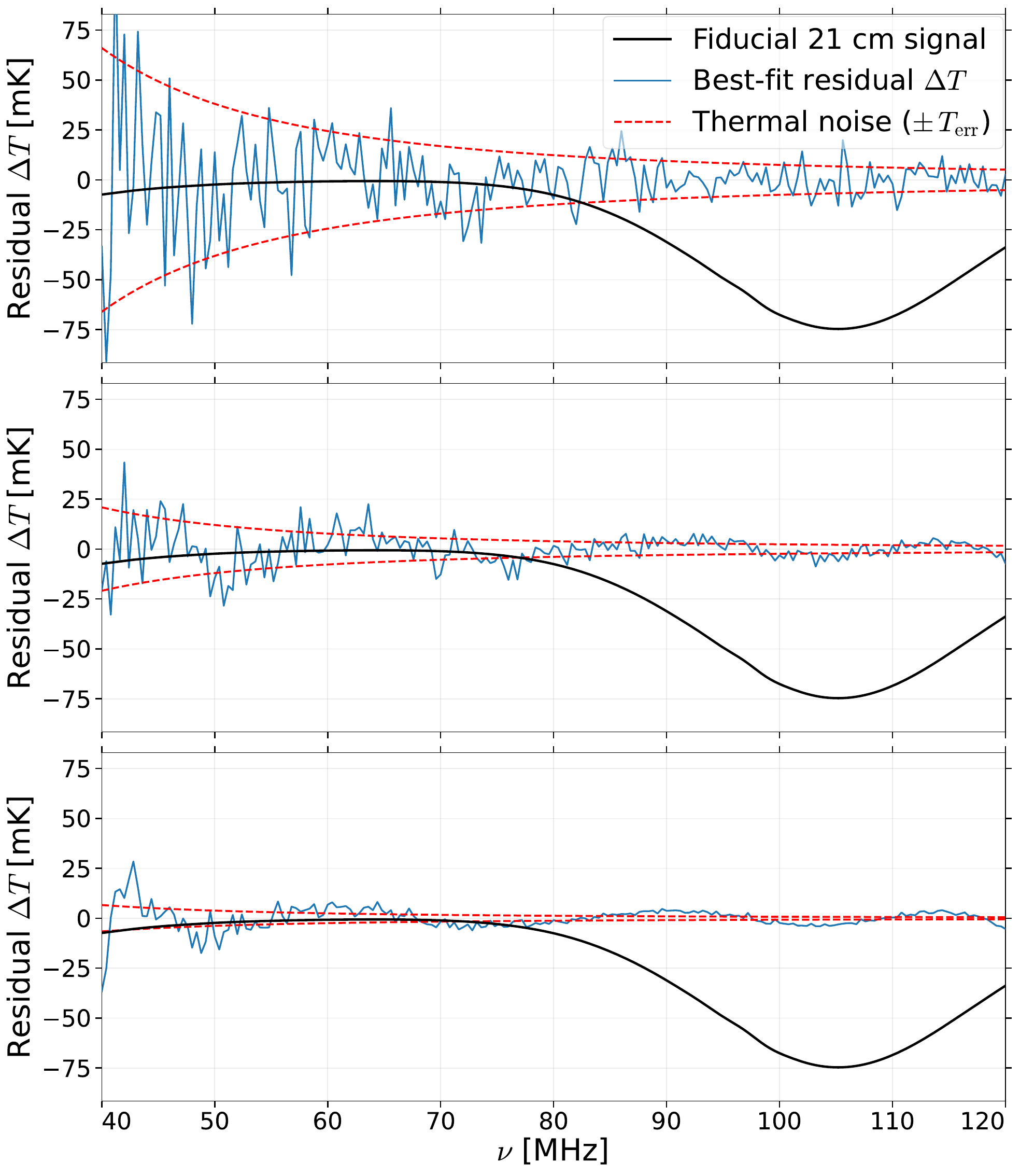}
\caption{Foreground-only fitting residuals for Model~1 with the broad priors in Table~\ref{tab:fg_prior_and_band}, for 30, 300, and 3,000 orbits from top to bottom. The blue curve is the residual of the median foreground-only fit, the other curves are similar to those in Figure~\ref{fig:global-spectrum-mock}(a). }
\label{fig:foreground_residue}
\end{figure}

For demonstration, we first consider a very narrow prior range for the foreground parameters which brackets their fiducial values.
In this case, one can obtain a fairly good fit to the 21\,cm global spectrum, which is shown in Figure~\ref{fig:global-spectrum-mock}, with panel~(a) for Model 1 and panel~(b) for Model 2. We plot the fiducial 21\,cm spectrum (black solid curve), the $\pm 1 \sigma$ thermal noise (red dashed lines) around the fiducial model, and the resulting channel-wise signal posterior (shades). The blue and orange shades are for the ideal dipole and CB-1 beams, respectively, but for the most part the two shades almost coincide with each other. The differences are mostly at the edges of the shades.

The thermal noise decreases with increasing frequency, as it is proportional to the system temperature, which is dominated by the sky temperature that scales roughly as $\nu^{-2.5}$. At $\nu < 60.4$ MHz, the posterior region is governed mainly by the signal prior, due to the higher system temperature and noise at lower frequencies. Between this frequency and 80\,MHz, the posterior uncertainty is generally consistent with the thermal-noise level.
Above 80\,MHz, however, the posterior uncertainty increases again and exceeds the thermal-noise level. This is due largely to the way we parameterize the foreground: we model the foreground as a polynomial expansion around $\nu =80$ MHz, and the foreground parameters in our model are those polynomial coefficients. Thus, at higher frequencies there is more leverage for the foreground, resulting in a larger uncertainty.
The posterior band narrows as the integration increases from 30 to 300 and 3,000 orbits.

Compared with Model 1, the absorption trough of Model 2 is deeper, and it is centered at a lower frequency (94 MHz as compared to 105 MHz in Model 1), so that nearly the whole frequency range of the absorption feature falls within the observational frequency band, while for Model 1 only part of it is within the observational frequency band.
For Model 2, the trough minimum is already well recovered at 30 orbits, while for Model 1, about 300 orbits are required to determine the trough minimum. With 3,000 orbits, the absorption minima are tightly reconstructed for both models, together with the transition into emission for Model~2.
Furthermore, for both models, the dipole and CB-1 posterior envelopes almost coincide, demonstrating that chromatic distortions are well absorbed by the foreground model when tight priors are available.

\begin{table}[th!]
\centering
\caption{The Fiducial Values and Priors of Foreground Coefficients. All coefficient values and ranges are in units of Kelvin.
The Tight Prior column gives representative ranges for the channel-wise joint fits in Figure~\ref{fig:global-spectrum-mock}(a), while the Broad Prior ranges are common to both models and beam configurations.
}
\label{tab:fg_prior_and_band}
\footnotesize
\setlength{\tabcolsep}{2pt}
\begin{tabular}{llll}
\toprule
 & Fiducial Value & Tight Prior & Broad Prior \\
\midrule
$a_0$ & $2164.6635$ & $[2164.6600,2164.6670]$ &  $[100,5000]$ \\
$a_1$ & $-55.3861$ & $[-55.417,-55.355]$ & $[-200,200]$ \\
$a_2$ & $-10.6874$ &  $[-10.800,-10.575]$ &  $[-200,200]$ \\
$a_3$ & $7.7869$ &  $[7.185,8.389]$ &  $[-50,50]$  \\
$a_4$ & $1.6143$ &  $[0.809,2.420]$ &  $[-50,50]$ \\
$a_5$ & $-1.5724$ &  $[-4.416,1.272]$ & $[-50,50]$ \\
$a_6$ & $0.2711$ &  $[-3.027,3.569]$ &  $[-50,50]$ \\
\bottomrule
\end{tabular}
\end{table}

Unfortunately, the successful fit achieved in this model depends critically on a relatively narrow prior on foreground parameters. The foreground parameters are allowed to vary within
approximately $\pm20 ~\sigma$ of their fiducial values, in which $\sigma$ denotes the posterior standard deviation of the
corresponding foreground coefficient, obtained by fitting the foreground model to the mock data with only foreground and thermal noise using MCMC.
Representative tight-prior ranges are listed in Table~\ref{tab:fg_prior_and_band}; their numerical values vary with the model and beam, whereas the broad-prior ranges remain unchanged.

If we remove the restriction on the foreground parameters and allow them to vary in a much broader range (the ``Broad Prior'' column in Table~\ref{tab:fg_prior_and_band}), then due to the degeneracy between the foreground and the 21\,cm signal, one cannot really separate these two components. Indeed, with a broad foreground prior, we can even fit the same mock data without any 21\,cm signal. The residual for the foreground-only fit in this case is shown in Figure~\ref{fig:foreground_residue}. We see that for the wider range of foreground parameters (coefficients), the mock data can be fit with the foreground model with very small residuals, without introducing any 21\,cm signal in the fit. If we do introduce the 21\,cm signal in the fit, this foreground solution would yield a 21\,cm signal very different from the actual one.
There would also be many other combinations of the 21\,cm signal and the foreground, so the posterior would be very broad.

This degeneracy between the 21\,cm signal and foreground in the global spectrum is fundamental. In order to break this degeneracy, one must either make assumptions about or place constraints on the foreground, restrict the form of the 21\,cm signal, or perhaps do both. For example, one might restrict the form of foreground by invoking the physical mechanism that generates it. In this paper, however, we shall allow the general foreground form, and instead restrict the 21\,cm signal to spectra generated by the \texttt{21cmEMU} model.
Below, we will investigate the constraints obtained with the 21~cm global spectrum alone, or by combining it with 21~cm power spectrum data.

\subsection{Global Spectrum Constraints on the {\tt 21cmEMU} Model Parameters}
 As discussed earlier, the \texttt{21cmEMU} model has nine parameters, but only five significantly affect the 21\,cm spectrum. Varying these parameters produces different spectra. Admittedly, this model has limitations: it assumes the standard cosmological model, does not consider new physics, and may omit important astrophysical processes that could significantly affect the 21\,cm signal. Nevertheless, within this framework, it should be possible to derive observational constraints on the model parameters from the global-spectrum measurement.

\begin{figure*}[tp]
\centering
\begin{minipage}[t]{0.47\textwidth}
\vspace{0pt}\centering
\includegraphics[width=\linewidth]{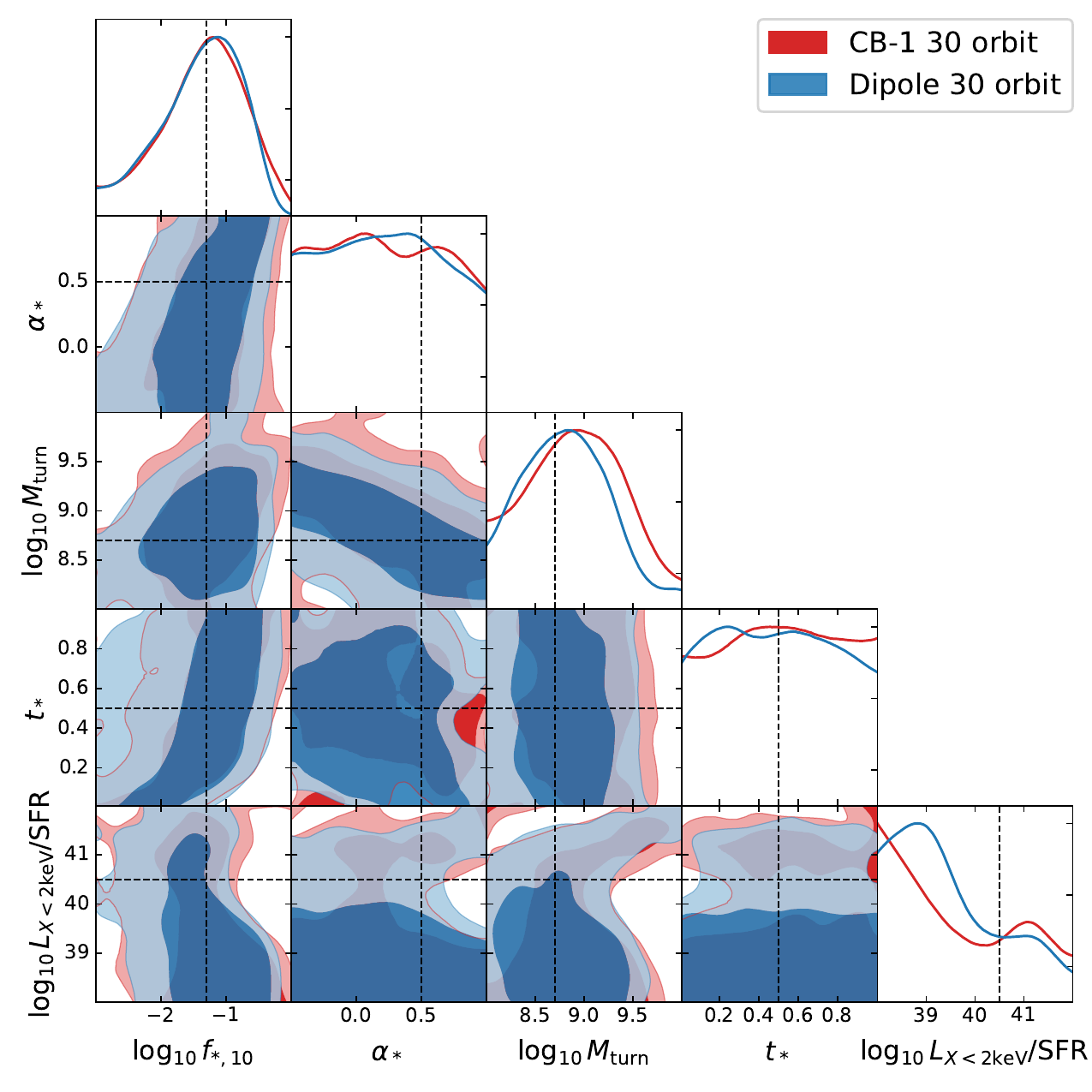}
\par\smallskip{\footnotesize {(a) Model 1, 30 orbits}}
\end{minipage}
\hfill
\begin{minipage}[t]{0.47\textwidth}
\vspace{0pt}\centering
\includegraphics[width=\linewidth]{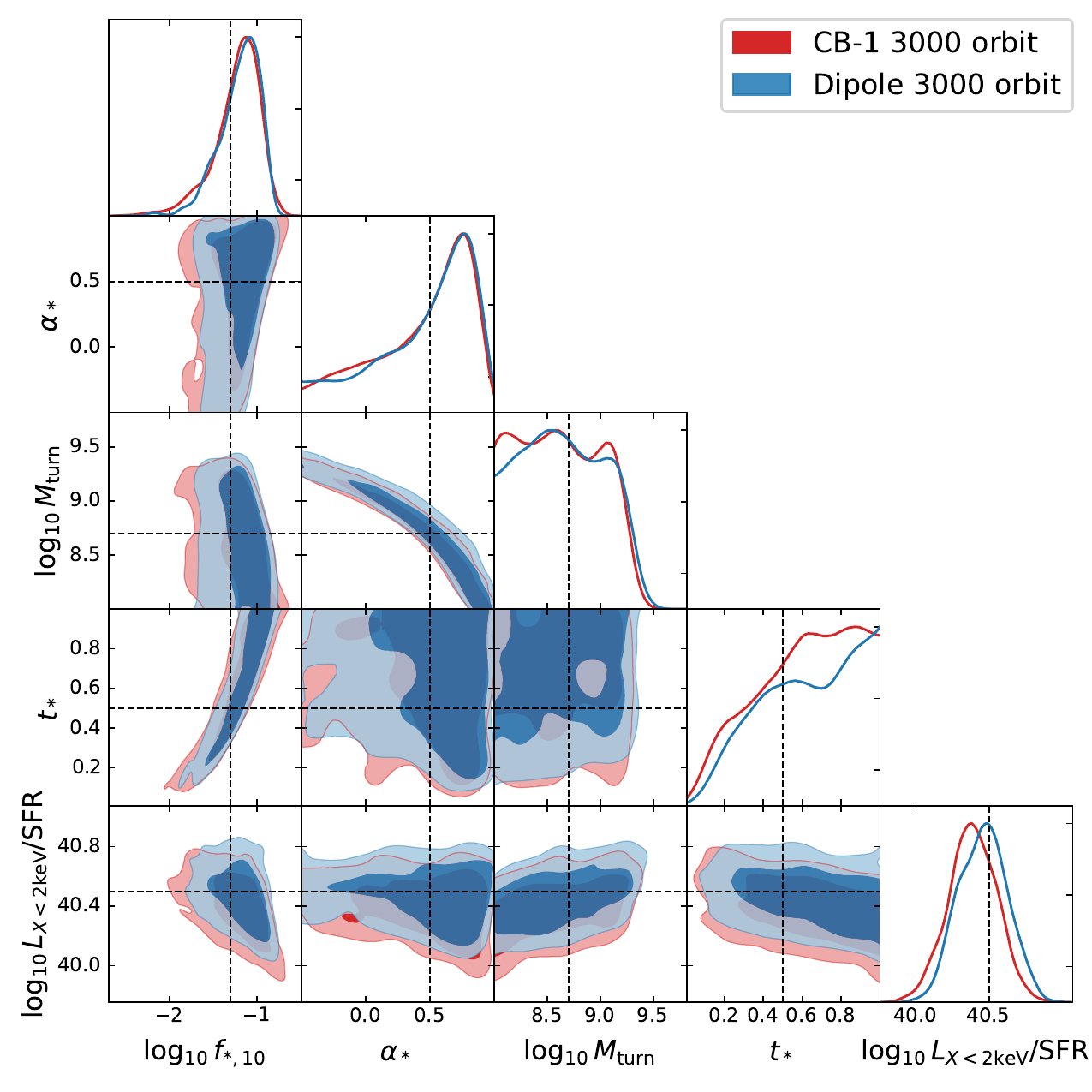}
\par\smallskip{\footnotesize {(b) Model 1, 3,000 orbits}}
\end{minipage}
\par\medskip
\begin{minipage}[t]{0.47\textwidth}
\vspace{0pt}\centering
\includegraphics[width=\linewidth]{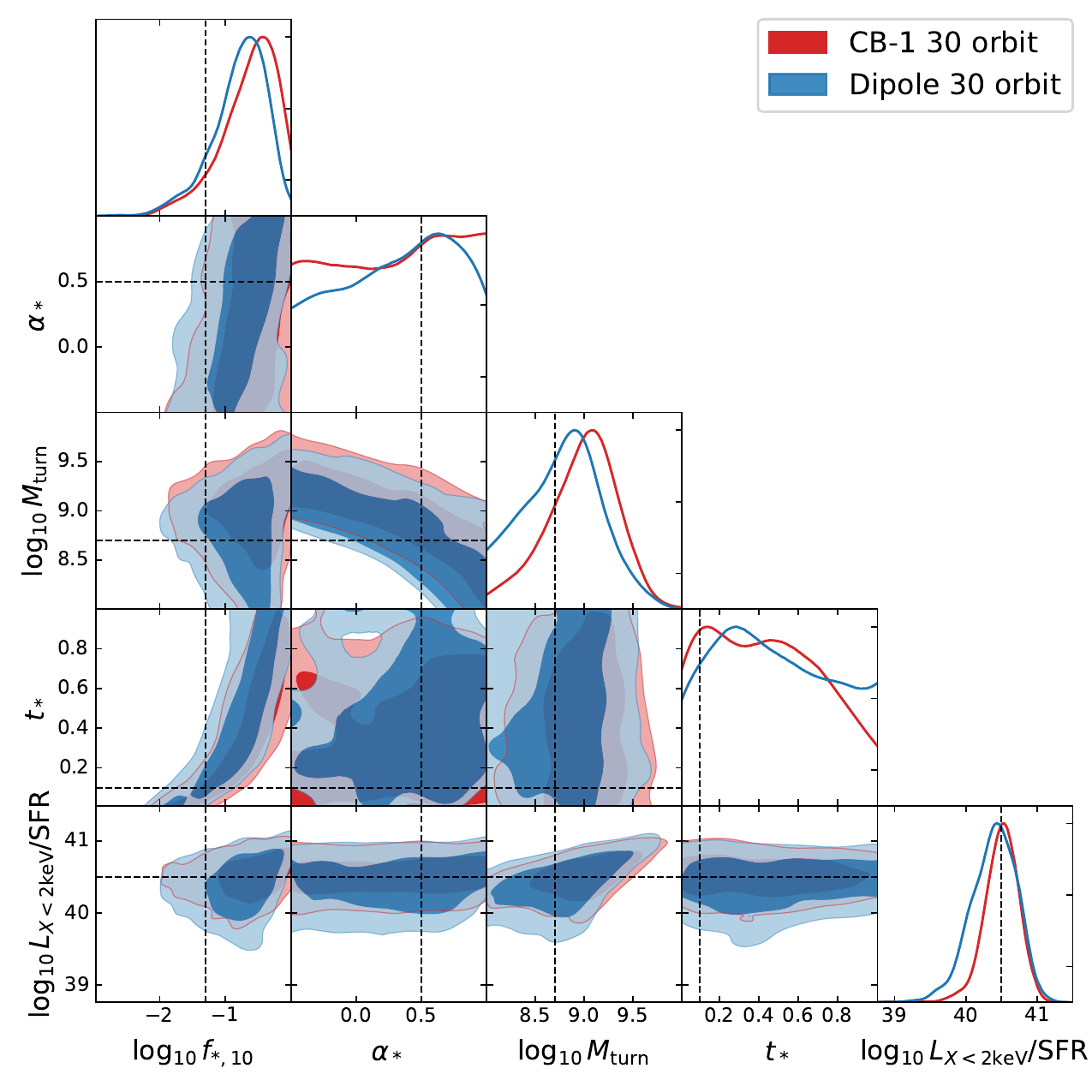}
\par\smallskip{\footnotesize {(c) Model 2, 30 orbits}}
\end{minipage}
\hfill
\begin{minipage}[t]{0.47\textwidth}
\vspace{0pt}\centering
\includegraphics[width=\linewidth]{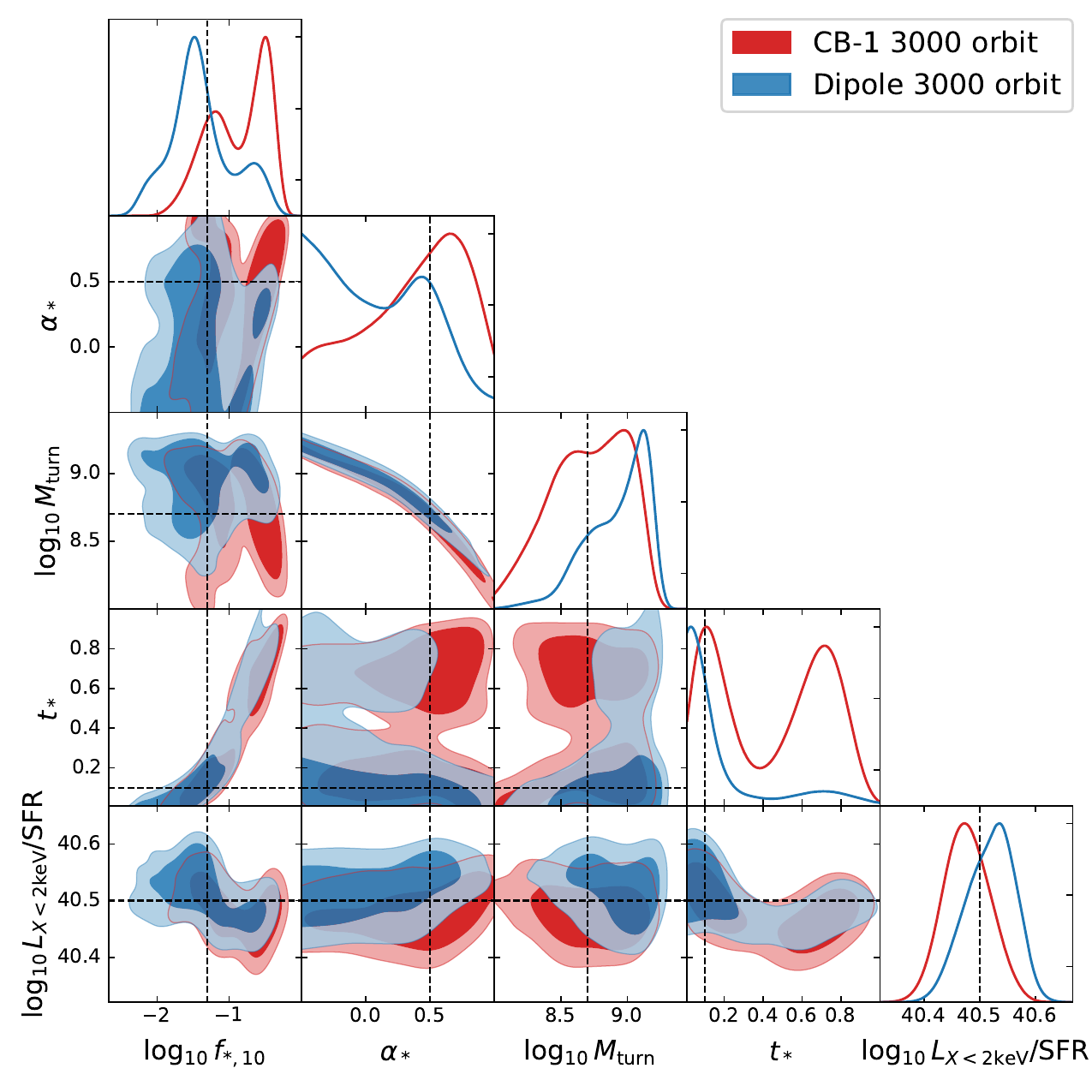}
\par\smallskip{\footnotesize {(d) Model 2, 3,000 orbits}}
\end{minipage}
\caption{Global-spectrum constraints on the five astrophysical parameters for Model 1 and Model 2, with foreground coefficients marginalized. The upper row shows the results for Model~1 and the lower row is for Model~2, at 30 orbits (left) and 3,000 orbits (right), respectively. Blue shades denote constraints obtained with the ideal dipole beam and red shades are for the constraints with the CB-1 beam; dashed lines mark each model's fiducial values. }
\label{fig:astrocorner_model2}
\end{figure*}

\begin{table*}[tbp]
\centering
\caption{Astrophysical Parameter Constraints for Model~1 from the DSL Global Spectrum with Foreground Parameters Marginalized. The listed fit values are the medians and the 16th/84th-percentile uncertainties when the posterior is peaked.  \textit{unconstrained} denotes a posterior approximately equal to the prior; \textit{prior-truncated} denotes a posterior that reaches a {\tt 21cmEMU} prior boundary.
}\label{tab:astro_param_constraints_model1}
{\renewcommand{\baselinestretch}{1.0}
\footnotesize
\setlength{\tabcolsep}{1.5pt}
\renewcommand{\arraystretch}{1.10}
\begin{tabular*}{\textwidth}{@{\extracolsep{\fill}}lrrrrrrr@{}}
\toprule
Parameter & Fiducial & \multicolumn{3}{c}{Dipole Beam} & \multicolumn{3}{c}{CB-1 Beam} \\
\cmidrule(lr){3-5}\cmidrule(lr){6-8}
& & 30 orbits & 300 orbits & 3,000 orbits & 30 orbits & 300 orbits & 3,000 orbits \\
\midrule
$\log_{10} f_{*,10}$ & $-1.30$ &
$-1.32^{+0.56}_{-0.73}$ & $-1.06^{+0.24}_{-0.43}$ & $-1.16^{+0.19}_{-0.28}$ &
$-1.30^{+0.59}_{-0.74}$ & $-1.35^{+0.44}_{-0.52}$ & $-1.18^{+0.20}_{-0.28}$ \\
$\alpha_{*}$ & $0.50$ &
$\textit{unconstrained}$ & $\textit{unconstrained}$ & $0.58^{+0.25}_{-0.53}$ &
$\textit{unconstrained}$ & $\textit{unconstrained}$ & $0.56^{+0.25}_{-0.56}$ \\
$\log_{10} M_{\mathrm{turn}}$ & $8.70$ &
$\textit{prior-truncated}$ & $9.06^{+0.40}_{-0.51}$ & $8.65^{+0.46}_{-0.42}$ &
$\textit{prior-truncated}$ & $9.03^{+0.42}_{-0.46}$ & $8.63^{+0.46}_{-0.43}$ \\
$t_{*}$ & $0.50$ &
$\textit{unconstrained}$ & $\textit{unconstrained}$ & $\textit{prior-truncated}$ &
$\textit{unconstrained}$ & $\textit{unconstrained}$ & $\textit{prior-truncated}$ \\
$\log_{10}(L_{\rm X<2\,keV}/{\rm SFR})$ & $40.50$ &
$\textit{unconstrained}$ & $40.51^{+0.32}_{-0.35}$ & $40.46^{+0.17}_{-0.18}$ &
$\textit{unconstrained}$ & $40.77^{+0.41}_{-0.90}$ & $40.39^{+0.17}_{-0.16}$ \\
\bottomrule
\end{tabular*}
}
\vspace{3pt}
\end{table*}

\begin{table*}[tbp]
\centering
\caption{{ Astrophysical Parameter Constraints, Similar to Table \ref{tab:astro_param_constraints_model1}, but for Model 2. The listed fit values are the medians and the 16th/84th-percentile uncertainties when the posterior is peaked. \textit{unconstrained} denotes a posterior approximately equal to the prior; \textit{prior-truncated} denotes a posterior that reaches a {\tt 21cmEMU} prior boundary. Bracketed entries denote 68\% credible intervals for parameters with two-peak posteriors.}
}\label{tab:astro_param_constraints_model2}
{\renewcommand{\baselinestretch}{1.0}
\footnotesize
\setlength{\tabcolsep}{1.5pt}
\renewcommand{\arraystretch}{1.10}
\begin{tabular*}{\textwidth}{@{\extracolsep{\fill}}lrrrrrrr@{}}
\toprule
Parameter & Fiducial & \multicolumn{3}{c}{Dipole Beam} & \multicolumn{3}{c}{CB-1 Beam} \\
\cmidrule(lr){3-5}\cmidrule(lr){6-8}
& & 30 orbits & 300 orbits & 3,000 orbits & 30 orbits & 300 orbits & 3,000 orbits \\
\midrule
$\log_{10}f_{*,10}$ & $-1.30$ & $-0.73^{+0.34}_{-0.47}$ & $-0.90^{+0.30}_{-0.38}$ &  $-1.44^{+0.59}_{-0.34}$ & $\textit{prior-truncated}$ & $-0.70^{+0.30}_{-0.47}$ & $[-1.34,\,-0.44]$ \\
$\alpha_*$ & $0.50$ & $\textit{unconstrained}$ & $\textit{prior-truncated}$ &  $\textit{prior-truncated}$ & $\textit{unconstrained}$ & $\textit{unconstrained}$ & $\textit{unconstrained}$ \\
$\log_{10}M_{\rm turn}$ & $8.70$ & $\textit{prior-truncated}$ & $8.88^{+0.19}_{-0.37}$ & $8.99^{+0.18}_{-0.31}$ & $9.01^{+0.32}_{-0.40}$ & $8.89^{+0.25}_{-0.40}$ & $8.73^{+0.30}_{-0.33}$ \\
$t_*$ & $0.10$ & $\textit{unconstrained}$ & $\textit{prior-truncated}$ & $\textit{prior-truncated}$ & $\textit{unconstrained}$ & $\textit{unconstrained}$ & $[0.11,\,0.77]$ \\
$\log_{10}(L_{\rm X<2\,keV}/{\rm SFR})$ & $40.50$ & $40.41^{+0.30}_{-0.35}$ & $40.43^{+0.10}_{-0.10}$ &  $40.52^{+0.04}_{-0.05}$ & $40.51^{+0.22}_{-0.24}$ & $40.44^{+0.08}_{-0.09}$ & $40.48^{+0.04}_{-0.04}$ \\
\bottomrule
\end{tabular*}
}
\end{table*}

\begin{figure}[tbp]
\centering
\includegraphics[width=0.95\columnwidth]{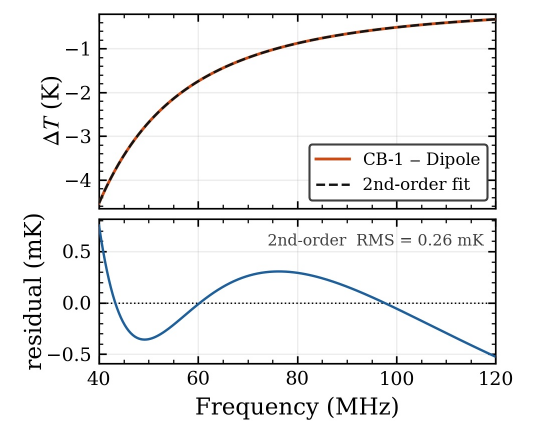}
\caption{The difference between the global spectra calculated using the dipole and CB-1 beams, together with the best-fit curve using a 2nd-order polynomial (upper panel). The fitting residual is shown in the lower panel, with an RMS of only 0.26\,mK.}
\label{fig:dp_minus_cbbeam}
\end{figure}

The numerical summaries of our results are given in Table~\ref{tab:astro_param_constraints_model1} for Model~1 and Table~\ref{tab:astro_param_constraints_model2} for Model~2. The allowed ranges of the parameters are given for the well-constrained ones. Parameters which are not well-constrained are marked as \textit{unconstrained} or \textit{prior-truncated} in the tables. \textit{unconstrained} denotes the cases where posteriors remain approximately equal to the prior, indicating that the data provide negligible information (the 95\% credible interval spans $\ge$90\% of the prior range); \textit{prior-truncated} denotes the cases where the posterior accumulates near a prior boundary, suggesting that the data favor values outside the imposed prior range but are restricted by the boundary.

\Needspace{5\baselineskip}
Figure~\ref{fig:astrocorner_model2} displays the posterior distributions for the five astrophysical parameters derived from the global spectrum for both Model~1 (panels~(a) and~(b)) and Model~2 (panels~(c) and~(d)), evaluated at 30 orbits (left column) and 3,000 orbits (right column). The ideal dipole and the CB-1 beam results are shown in shades of different colors.

For Model~1, at 30 orbits (Figure~\ref{fig:astrocorner_model2}(a)), most parameters are not well-constrained, but at 3,000 orbits (Figure~\ref{fig:astrocorner_model2}(b)) the allowed parameter space shrinks a lot, and we obtain very good constraints on $f_*$ and $L_X/\mathrm{SFR}$, and we see strong degeneracy between $\alpha_*$ and $M_{\rm turn}$, though there remains large uncertainty in $t_*$ and $M_{\rm turn}$.

The contours for the ideal dipole and CB-1 posterior only have moderate differences, the fiducial value is within the credible intervals of both. To understand this similarity of result for the two antenna beams, we plot the difference in the measured global spectrum by the ideal dipole and the CB-1 antenna in the top panel of Figure~\ref{fig:dp_minus_cbbeam}. Although the chromaticity and sky anisotropy induce a spectrum deviation of a few Kelvins in the observational frequency range, we see it is a smooth function of frequency, which can be adequately described by a 2nd-order polynomial. As we are already fitting the foreground with a seven-term polynomial, a large part of this chromaticity-induced spectral distortion will be absorbed into the foreground, with a residual RMS of 0.26\,mK, as shown in the bottom panel of Figure~\ref{fig:dp_minus_cbbeam}. This explains why the difference in the constrained parameters is small.

For Model~2, at 30 orbits (Figure~\ref{fig:astrocorner_model2}(c)), some parameters, particularly $L_X/\mathrm{SFR}$, are better constrained than for Model~1, possibly because the absorption feature is more pronounced and lies well within the observational band of the DSL high-frequency spectrometer. At 3,000 orbits (Figure~\ref{fig:astrocorner_model2}(d)), the constraints on $L_X/\mathrm{SFR}$ become much tighter, while substantial uncertainties remain in $f_*$, $\alpha_*$, and $t_*$. These remaining uncertainties likely reflect degeneracies among the astrophysical parameters and with the foreground parameters.

The difference between the dipole beam and the CB-1 beam is more apparent in Model 2 at 3,000 orbits. The spectrum deviation induced by the beam chromaticity should be almost the same as in the case of Model 1, but in terms of the inferred astrophysical parameter, the impact is larger.  While the input parameter values generally lie within the $1 \sigma$ contours of the ideal dipole, the contours for the CB-1 beam deviate significantly. This shows that the chromaticity of the beam could indeed induce large systematic errors.

\subsection{Joint Inference with the 21 cm Power Spectrum}\label{subsec:joint_inference}

The global signal and the 21\,cm power spectrum constrain complementary aspects of the same astrophysical model: the mean thermal and ionization history and the spatial fluctuations, respectively. We combine the 3,000-orbit CB-1 global-spectrum posterior with the HERA Moderate and SKA-Low AA4 Moderate deep-survey power-spectrum posteriors (evaluated under the observing configurations described in Table~\ref{tab:psonly_obs_params}). The combined constraints are obtained by post-processing the separately sampled global-spectrum and power-spectrum posteriors, as described in Section~\ref{sec:methodology}.
The corresponding power-spectrum constraints based on \texttt{21cmEMU} \citep{Breitman2023} are obtained with forecast uncertainties from \texttt{21cmSense} \citep{pober2014next}.

\begin{figure*}[tp]
\centering
\begin{minipage}[t]{0.47\textwidth}
\vspace{0pt}\centering
\includegraphics[width=\linewidth]{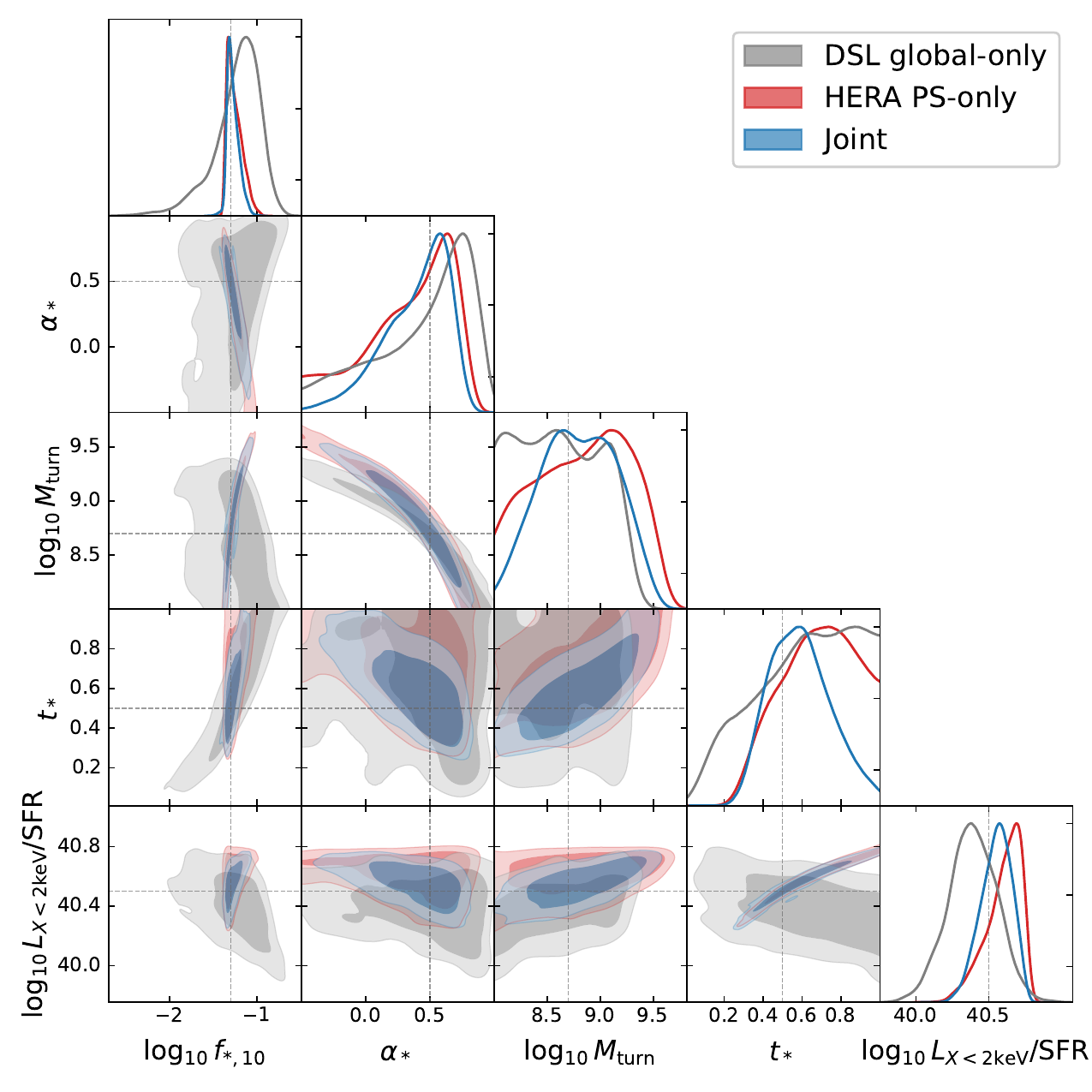}
\par\smallskip{\footnotesize {(a) Model 1, HERA}}
\end{minipage}
\hfill
\begin{minipage}[t]{0.47\textwidth}
\vspace{0pt}\centering
\includegraphics[width=\linewidth]{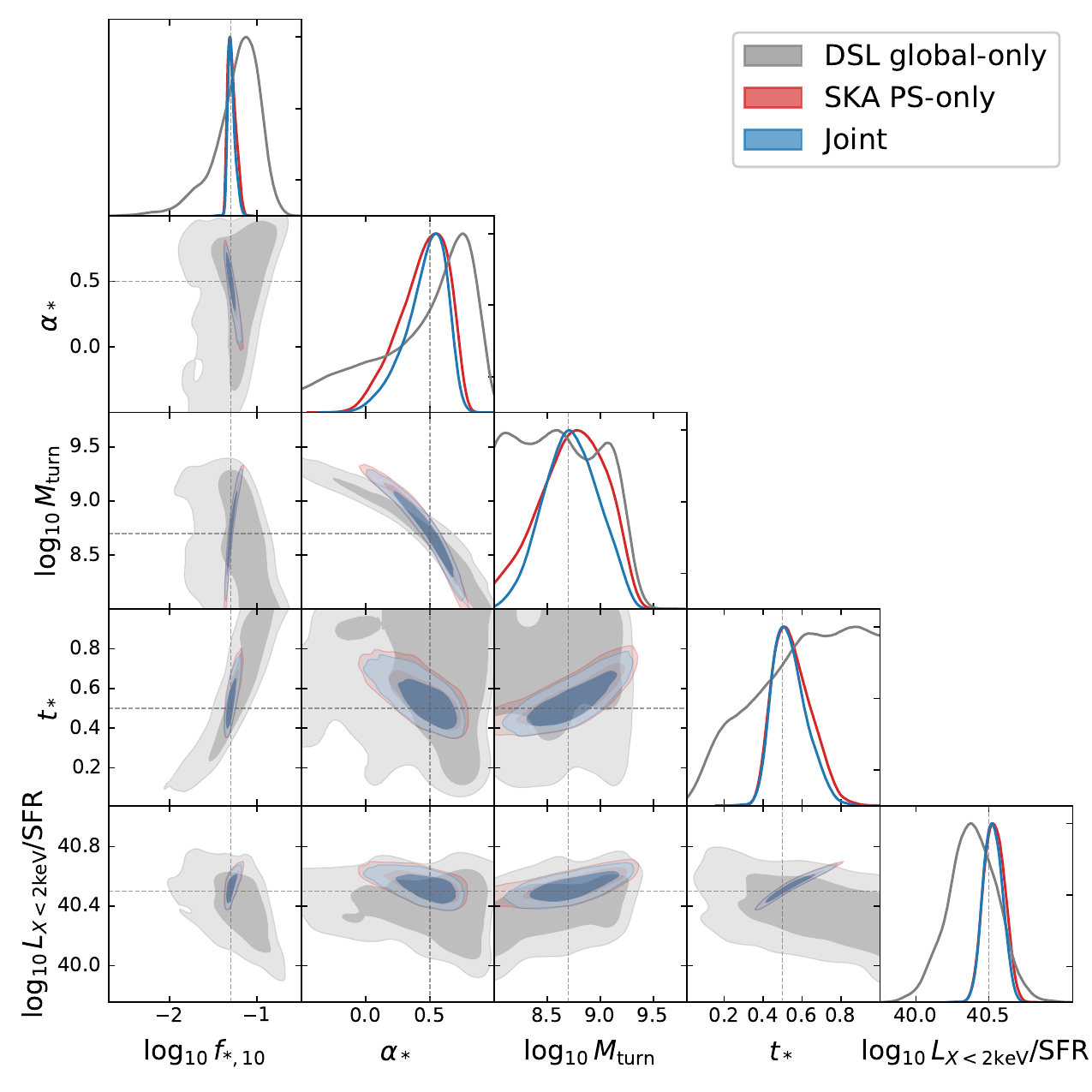}
\par\smallskip{\footnotesize {(b) Model 1, SKA}}
\end{minipage}
\par\medskip
\begin{minipage}[t]{0.47\textwidth}
\vspace{0pt}\centering
\includegraphics[width=\linewidth]{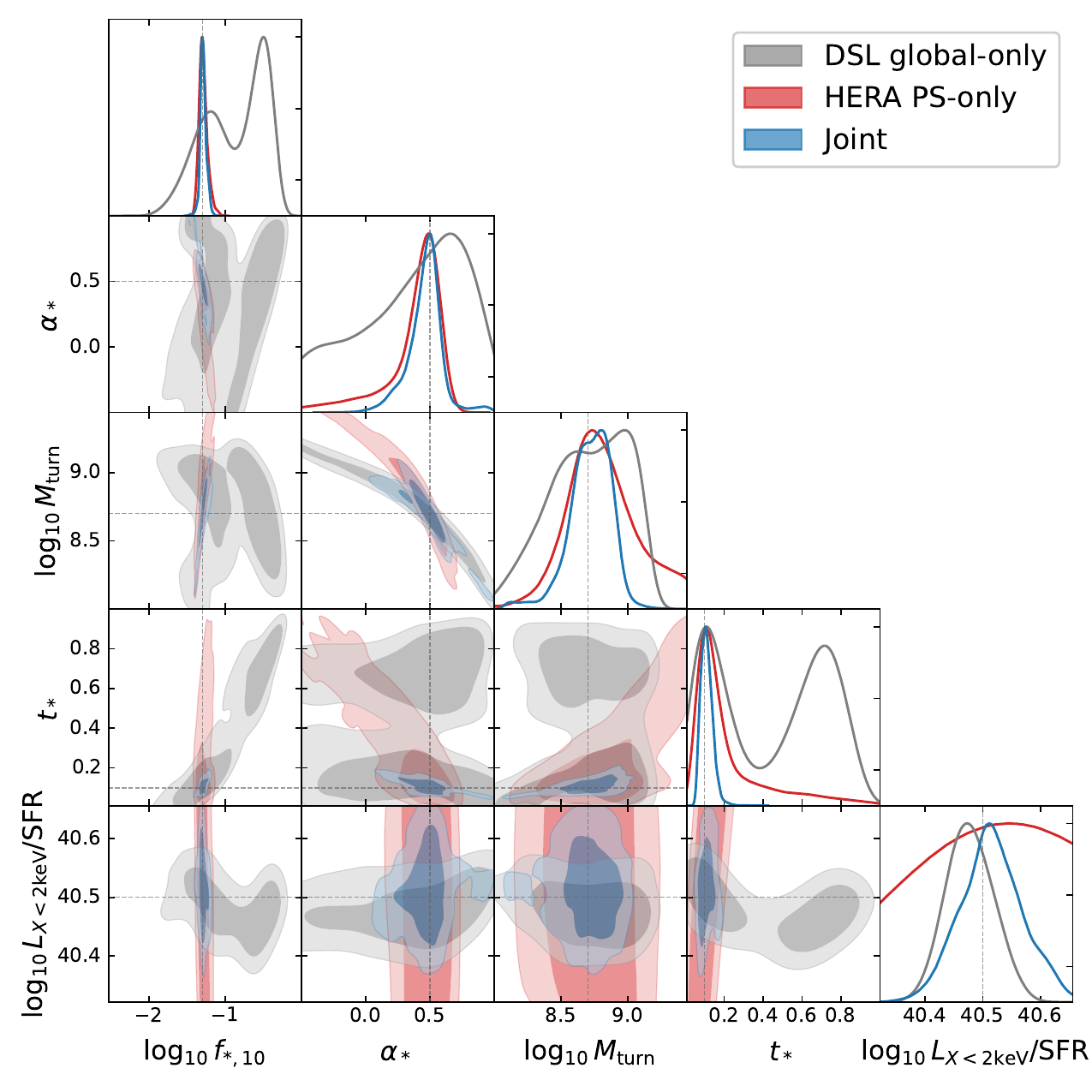}
\par\smallskip{\footnotesize {(c) Model 2, HERA}}
\end{minipage}
\hfill
\begin{minipage}[t]{0.47\textwidth}
\vspace{0pt}\centering
\includegraphics[width=\linewidth]{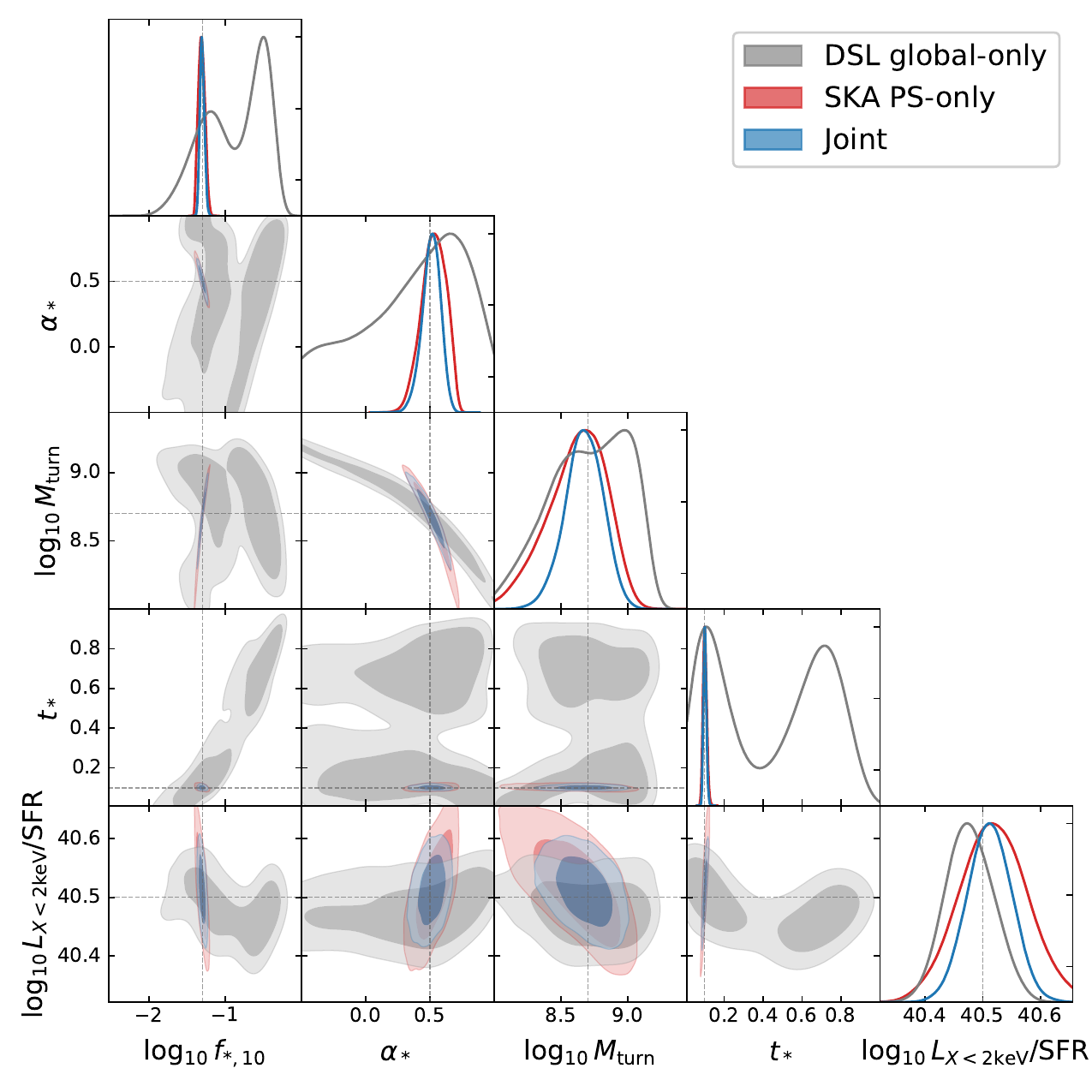}
\par\smallskip{\footnotesize {(d) Model 2, SKA}}
\end{minipage}
\caption{Joint posterior comparisons for Model~1 (upper row) and Model~2 (lower row). The DSL global-only component uses the 3,000-orbit CB-1 analysis. Grey, red, and blue shades denote DSL global-only, power-spectrum-only, and combined posteriors, respectively. The left column uses HERA Moderate and the right column uses SKA-Low AA4 Moderate in the deep-survey configuration. Dashed lines mark the respective fiducial values. Numerical intervals are given in Tables~\ref{tab:joint_combined_models} and \ref{tab:joint_combined_model2}.}
\label{fig:corner-comb-joint}
\end{figure*}

Figure~\ref{fig:corner-comb-joint} shows the constraints for Model~1 in combination with HERA in panel~(a), and in combination with SKA-Low in panel~(b); and the constraints for Model~2 in combination with HERA in panel~(c), and with SKA-Low in panel~(d). The corresponding numerical summaries are listed separately in Tables~\ref{tab:joint_combined_models} and \ref{tab:joint_combined_model2}.

As can be seen from the figures and tables,
the complementary information from the power spectrum  breaks the broad global-signal degeneracies, and the constraints are substantially improved. For all  recovered model parameters, the joint credible intervals are generally consistent with the input parameter values.

For Model~1, combining the DSL global spectrum with either HERA or SKA-Low tightens the constraints on the star-formation efficiency normalization and X-ray efficiency relative to the global-only analysis (Table~\ref{tab:joint_combined_models}). Even
for $M_{\rm turn}$ and $t_*$, which remain prior-truncated with the global spectrum only or with the HERA power spectrum only, the joint inference yields effective constraints.

For Model~2, the improvement over the global spectrum-only constraint is similar, but the power spectrum is especially informative because the signal itself is stronger: the shorter star-formation timescale $t_*=0.1$ makes the Ly$\alpha$ coupling start earlier,
producing larger 21~cm spatial fluctuations, while the array sensitivity is the same for both fiducial signals. Indeed, the HERA power spectrum data alone can already constrain $t_*$ and $M_{\rm turn}$ in Model~2. The DSL global-spectrum posterior provides complementary information, particularly for HERA; for example, adding DSL narrows the reported 68\% marginal credible interval for $t_*$ by approximately a factor of five relative to HERA alone (Table~\ref{tab:joint_combined_model2}).
Combining the 3,000-orbit DSL CB-1 observations with the SKA-Low power spectrum yields tight constraints on all five parameters, with the fiducial values lying within the reported 68\% marginal credible intervals (Table~\ref{tab:joint_combined_model2}).

\begin{table*}[tbp]
\centering
\caption{Constraints on the Five Astrophysical Parameters for Model~1 from the DSL CB-1 Global Spectrum at 3,000 Orbits and the HERA and SKA-Low AA4 Moderate Power-Spectrum Forecasts. The SKA forecast uses the deep-survey configuration. Entries give posterior medians with 68\% credible intervals when the posterior is peaked; \textit{prior-truncated} denotes a posterior that reaches a {\tt 21cmEMU} prior boundary. PS-only denotes a power-spectrum analysis alone; joint denotes its combination with DSL.}\label{tab:joint_combined_models}
{\renewcommand{\baselinestretch}{1.0}
\footnotesize
\setlength{\tabcolsep}{1.5pt}
\renewcommand{\arraystretch}{1.10}
\begin{tabular*}{\textwidth}{@{\extracolsep{\fill}}lrrrrrr@{}}
\toprule
Parameter & Fid. & DSL CB-1 & HERA PS-only & SKA PS-only & DSL+HERA & DSL+SKA \\
& & 3,000 orbits & Moderate &  Moderate & Joint & Joint \\
\midrule
$\log_{10} f_{*,10}$ & $-1.30$ & $-1.18^{+0.20}_{-0.28}$ & $-1.27^{+0.11}_{-0.07}$ & $-1.29^{+0.06}_{-0.04}$ &  $-1.29^{+0.08}_{-0.05}$ &  $-1.30^{+0.05}_{-0.03}$ \\
$\alpha_{*}$ & $0.50$ & $0.56^{+0.25}_{-0.56}$ & $0.42^{+0.27}_{-0.44}$ &  $0.47^{+0.17}_{-0.23}$ &  $0.45^{+0.20}_{-0.32}$ & $0.49^{+0.13}_{-0.19}$ \\
$\log_{10} M_{\mathrm{turn}}$ & $8.70$ & $\textit{prior-truncated}$ & $\textit{prior-truncated}$ &  $8.76^{+0.31}_{-0.35}$ &  $8.80^{+0.38}_{-0.37}$ &  $8.72^{+0.28}_{-0.26}$ \\
$t_{*}$ & $0.50$ & $\textit{prior-truncated}$ & $\textit{prior-truncated}$ &  $0.54^{+0.12}_{-0.09}$ &  $0.58^{+0.17}_{-0.15}$ &  $0.53^{+0.10}_{-0.08}$ \\
$\log_{10}(L_{\rm X<2\,keV}/{\rm SFR})$ & $40.50$ & $40.39^{+0.17}_{-0.16}$ & $40.63^{+0.10}_{-0.15}$ &  $40.54^{+0.07}_{-0.07}$ & $40.56^{+0.10}_{-0.12}$ &  $40.53^{+0.07}_{-0.06}$ \\
\bottomrule
\end{tabular*}
}
\end{table*}

\begin{table*}[tbp]
\centering
\caption{{ Constraints on the Five Astrophysical Parameters, Similar to Table \ref{tab:joint_combined_models} but for Model 2. The bracketed entry denotes 68\% credible interval for the parameters with a two-peak posterior.}}\label{tab:joint_combined_model2}
{\renewcommand{\baselinestretch}{1.0}
\footnotesize
\setlength{\tabcolsep}{1.5pt}
\renewcommand{\arraystretch}{1.10}
\begin{tabular*}{\textwidth}{@{\extracolsep{\fill}}lrrrrrr@{}}
\toprule
Parameter & Fid. & DSL CB-1 & HERA PS-only & SKA PS-only & DSL+HERA & DSL+SKA \\
& & 3,000 orbits & Moderate & Moderate & Joint & Joint \\
\midrule
$\log_{10}f_{*,10}$ &  $-1.30$ & $[-1.34,\,-0.44]$ &
$-1.30^{+0.06}_{-0.05}$ &
$-1.31^{+0.05}_{-0.04}$ &
$-1.29^{+0.04}_{-0.03}$ &
$-1.31^{+0.03}_{-0.03}$ \\
$\alpha_*$ &  $0.50$ &
$\textit{unconstrained}$ &
$0.45^{+0.11}_{-0.24}$ &
$0.53^{+0.10}_{-0.10}$ &
$0.48^{+0.09}_{-0.14}$ &
$0.51^{+0.07}_{-0.07}$ \\
$\log_{10}M_{\rm turn}$ & $8.70$ &
$8.73^{+0.30}_{-0.33}$ &
$8.79^{+0.31}_{-0.22}$ &
$8.64^{+0.20}_{-0.25}$ &
$8.73^{+0.13}_{-0.15}$ &
$8.67^{+0.14}_{-0.14}$ \\
$t_*$ & $0.10$ &
$[0.11,\,0.77]$&
$0.13^{+0.25}_{-0.05}$ &
$0.10^{+0.01}_{-0.01}$ &
$0.11^{+0.03}_{-0.03}$ &
$0.10^{+0.01}_{-0.01}$ \\
$\log_{10}(L_{\rm X<2\,keV}/{\rm SFR})$ &
$40.50$ & $40.48^{+0.04}_{-0.04}$ & $40.66^{+0.56}_{-0.26}$&
$40.52^{+0.06}_{-0.06}$ &
$40.51^{+0.06}_{-0.05}$ &
$40.51^{+0.04}_{-0.04}$ \\
\bottomrule
\end{tabular*}
}
\end{table*}

\section{Conclusion}\label{sec:conclusion}
 We have developed a Bayesian framework to forecast the astrophysical constraints accessible to the lunar-orbit DSL global 21\,cm experiment. The analysis combines \texttt{21cmEMU} predictions with foreground emission, orbital averaging, lunar occultation, beam chromaticity, and thermal noise across two representative scenarios: Model~1 ($t_*=0.5$) and Model~2 ($t_*=0.1$ with accelerated Cosmic Dawn). We model the foreground over 40--120\,MHz with a seven-term logarithmic polynomial, and obtain the residue errors assuming
 the thermal noises for a total observing time of 30, 300, and 3,000 orbits (corresponding to far-side integration times of about 23, 230, and 2300\,h, respectively).

 The channel-wise recovery analysis illustrates the degeneracy between the foreground and the cosmological signal. With tight foreground priors (cf. Table~\ref{tab:fg_prior_and_band}), the 21~cm signal can be recovered. With broad foreground priors, however, there is enough flexibility in the foreground model that even a foreground-only model also fits the mock spectrum with small residuals.  Thus, precise measurements of the total spectrum and small fitting residuals do not by themselves establish that the 21~cm signal has been correctly separated from the foreground. To break this degeneracy between the 21~cm signal and foreground, one has to make some further assumptions on either the 21~cm signal or the foreground, or both.

In this work we allow the foreground model to be general while restricting the 21~cm signal to \texttt{21cmEMU} models, which allows constraining the astrophysical parameters which determine the 21~cm signal. In the global-only analysis of Model~1, $\log_{10}f_{*,10}$ is already constrained at 30 orbits, and the X-ray efficiency becomes constrained from 300 orbits onward. $\alpha_*$ is degenerate with $M_{\rm turn}$, while $t_*$ is either unconstrained or the posterior is prior-truncated, only for Model 2 at 3,000 orbits can some broad constraint be placed on it.

We also study the impact of chromatic beam on the inference result by comparing an ideal achromatic dipole beam with the mildly chromatic beam CB-1 produced by an ice cream antenna. The beam chromaticity coupled with sky anisotropy induces a deviation in the measured spectrum, but this deviation can be well-fit by a low-order polynomial and thus absorbed in the foreground fit. If the precision of the measurement is not high, then the effect is insignificant. However, in highly precise measurement, e.g. the 3,000 orbit with Model 2, it can bias the inference result.

Finally we study the synergy of the 21~cm global spectrum and the 21~cm spatial power spectrum. For the power spectrum, we consider the HERA and SKA-Low, adopting the ``Moderate'' foreground-avoidance prescription \citep{pober2013,pober2014next}. The joint analysis of the global spectrum and power-spectrum completely breaks the degeneracy and yields  powerful constraints. For Model~1, while the DSL-only and the HERA-only posteriors for $M_{\rm turn}$ and $t_*$ remain prior-truncated, the combination yields well-constrained posterior distributions for all five parameters.
Also, for Model 2, the DSL-only posterior for $t_*$ remains poorly-constrained, while the HERA-only posterior for $L_X/\mathrm{SFR}$ is poorly-constrained, much better constraints can be achieved in the joint analysis thanks to the complementary information.

These results show that high precision 21~cm global spectrum data, such as that of the DSL mission, can be a powerful probe of the astrophysical processes of the cosmic dawn and epoch of reionization, but there are also great challenges in extracting the 21~cm signal from the strong foreground, and inferring the astrophysical parameters from the data.  The 21~cm signal extraction may be improved by taking into account the anisotropy of the foreground and time variation of the data \citep{Li2025}, and by refining the foreground model by considering its physical mechanism. On the other hand, the beam of the DSL may have more complicated form than the CB-1 beam adopted here, once more detailed design requirements are taken into account. We shall investigate these issues further in subsequent works.

\begin{acknowledgements}

We gratefully acknowledge Fengquan Wu and his team, esp. Jiajun Yuan for their contributions to the beam simulations and for providing the resulting antenna profiles. The computations were carried out on the Tianhe new generation supercomputer at the National Supercomputer Center in Tianjin. This work is supported by the Space Origins Exploration Program Nos. GJ11010405, GJ11010401, National Key R\&D Program of China No. 2022YFF0504300, and the National Natural Science Foundation of China (NSFC) grant Nos. 12361141814, 12533001, 12421003, 12273070.
\end{acknowledgements}

\noindent\textbf{Facilities:} DSL.\par

\noindent\textbf{Software:} 21cmEMU \citep{Breitman2023}, emcee \citep{emcee}.\par

\FloatBarrier
\bibliographystyle{raa}
\bibliography{reference}

\end{document}